\documentclass[letterpaper]{article} 
\usepackage{aaai25}  
\usepackage{times}  
\usepackage{helvet}  
\usepackage{courier}  
\usepackage[hyphens]{url}  
\usepackage{graphicx} 
\usepackage{natbib}  
\usepackage{caption} 
\usepackage{algorithm}
\usepackage{algpseudocode}
\usepackage{booktabs}
\usepackage{multirow}
\usepackage{enumitem}
\usepackage{amssymb} 

\usepackage{newfloat}
\usepackage{listings}
\DeclareCaptionStyle{ruled}{labelfont=normalfont,labelsep=colon,strut=off} 
\floatstyle{ruled}
\newfloat{listing}{tb}{lst}{}
\floatname{listing}{Listing}
\title{Multi-granularity Score-based Generative Framework Enables Efficient Inverse Design of Complex Organics}
\author{
    Zijun Chen,\textsuperscript{\rm 1,}\equalcontrib \
    Yu Wang,\textsuperscript{\rm 1,}\equalcontrib \
    Liuzhenghao Lv\textsuperscript{\rm 1},
    Hao Li\textsuperscript{\rm 1,}\textsuperscript{\rm 2}, 
    Zongying Lin\textsuperscript{\rm 1},
    Li Yuan\textsuperscript{\rm 1,}\textsuperscript{\rm 2,}\thanks{Corresponding author: Li Yuan (yuanli-ece@pku.edu.cn)},
    Yonghong Tian\textsuperscript{\rm 1,}\textsuperscript{\rm 2,}\thanks{Corresponding author: Yonghong Tian (yhtian@pku.edu.cn)}
}
\affiliations{
    \textsuperscript{\rm 1} Peking University, China
    \textsuperscript{\rm 2} Peng Cheng Laboratory, China

}

\usepackage{bibentry}
\usepackage{amsfonts}
\usepackage{amsmath}
\usepackage{textgreek}

\begin{document}

\maketitle

\begin{abstract}
Efficiently retrieving an enormous chemical library to design targeted molecules is crucial for accelerating drug discovery, organic chemistry, and optoelectronic materials. Despite the emergence of generative models to produce novel drug-like molecules, in a more realistic scenario, the complexity of functional groups (e.g., pyrene, acenaphthylene, and bridged-ring systems) and extensive molecular scaffolds remain challenging obstacles for the generation of complex organics. Traditionally, the former demands an extra learning process, e.g., molecular pre-training, and the latter requires expensive computational resources. To address these challenges, we propose OrgMol-Design, a multi-granularity framework for efficiently designing complex organics. Our OrgMol-Design is composed of a score-based generative model via fragment prior for diverse coarse-grained scaffold generation and a chemical-rule-aware scoring model for fine-grained molecular structure design, circumventing the difficulty of intricate substructure learning without losing connection details among fragments. Our approach achieves state-of-the-art performance in four real-world and more challenging benchmarks covering broader scientific domains, outperforming advanced molecule generative models. Additionally, it delivers a substantial speedup and graphics memory reduction compared to diffusion-based graph models. Our results also demonstrate the importance of leveraging fragment prior for a generalized molecule inverse design model.
\end{abstract}

\section{Introduction}

Molecule inverse design is to produce molecular structures with desired properties, which is deemed as the holy grail in material science ~\cite{weiss2023guided, takeda2020molecular}, organic chemistry ~\cite{nigam2024artificial}, and biomedical drug discovery ~\cite{igashov2024equivariant, swanson2024generative, jiang2024pocketflow}. Traditional protocols to devise novel compounds with specific demands are mainly based on domain experts ~\cite{molesky2018inverse, grigalunas2021natural}, in-silicon screening ~\cite{lyu2023modeling}, and high throughput wet-lab experiments ~\cite{zeng2023high}, involving with frequent searching in the exploding atom-level combination space to uncover latent quantitative structure-activity relationships ~\cite{tropsha2024integrating}. 

With the accumulation of available molecular data and the proliferation of computational resources, deep generative models are emerging as promising approaches for designing novel molecules. These deep generative methods typically employ atom-level descriptions of 3-dimension (3D) spaces. Previous generation models, such as diffusion-based models ~\cite{hoogeboom2022equivariant, huang2023mdm, xu2024geometric}, flow-matching-based models ~\cite{dunn2024mixed, song2024equivariant} and Bayesian-flow-networks-based models \cite{song2023unified, qu2024molcraft}, concentrated on the generation molecules in 3D space, given that the spatial information of 3D molecules can be conveniently described at the atomic level in continuous space (e.g., using Cartesian coordinates of atoms). However, in the design of complex organic compounds, generating 3D conformations using generative models is often both challenging and unnecessary \cite{zheng2024structure}. This difficulty arises from the fact that accurately computing the conformations of millions of complex molecules to train the generative model is not feasible in a timely manner. Furthermore, these conformations are dynamic within the system and are influenced by multiple environmental factors. Using deep generative models to produce merely a single snapshot in real space, rather than a distribution, is often insufficient. Even in some works such as DiG \cite{zheng2024predicting}, where these distributions can be obtained, researchers frequently need to employ computational chemistry methods to optimise these conformations during the practical application of the models.

Fortunately, several discrete diffusion generative methods based on molecular graphs have already been proposed. For instance, EDP-GNN \cite{niu2020permutation} modeled gradients of input graph with permutation equivariant graph networks. Similarly, GraphGDP \cite{huang2022graphgdp} and DiGress\cite{vignac2022digress} devised a position-enhanced graph score network and a discrete denoising diffusion process, respectively. However, these methods typically require frequent operations on an adjacency matrix, whose complexity scales quadratically with the number of nodes in the graph. This significantly limits the scalability of diffusion-based molecular generation methods to larger molecular scaffolds. Moreover, as the number of atoms increases, the combinatorial space of complex molecules grows exponentially. However, chemically valid topological structures lie on a low-dimensional manifold within this high-dimensional combinatorial space. During the diffusion process, noisy structures often end up outside this manifold, further increasing the learning cost for diffusion generative models.

To address the aforementioned challenges, methods that use molecular fragments as basic token units for molecular generation have been proposed and have gradually achieved impressive performance via variational auto-encoders (VAE) \cite{jin2018junction, jin2020hierarchical, geng2023de, chen2021deep}. This is because using molecular fragments as the smallest descriptive units in molecule generation can significantly reduce the number of nodes in the molecular graph. Additionally, due to the inherent prior knowledge of molecular structures within the fragments, even the random combination of different molecular fragments \cite{wu2024t} or customized sampling strategies \cite{xie2021mars, fu2021mimosa, guo2022data} can yield desirable performance on benchmark datasets. 

\begin{figure}[h]
    \centering
    \includegraphics[width=1\linewidth]{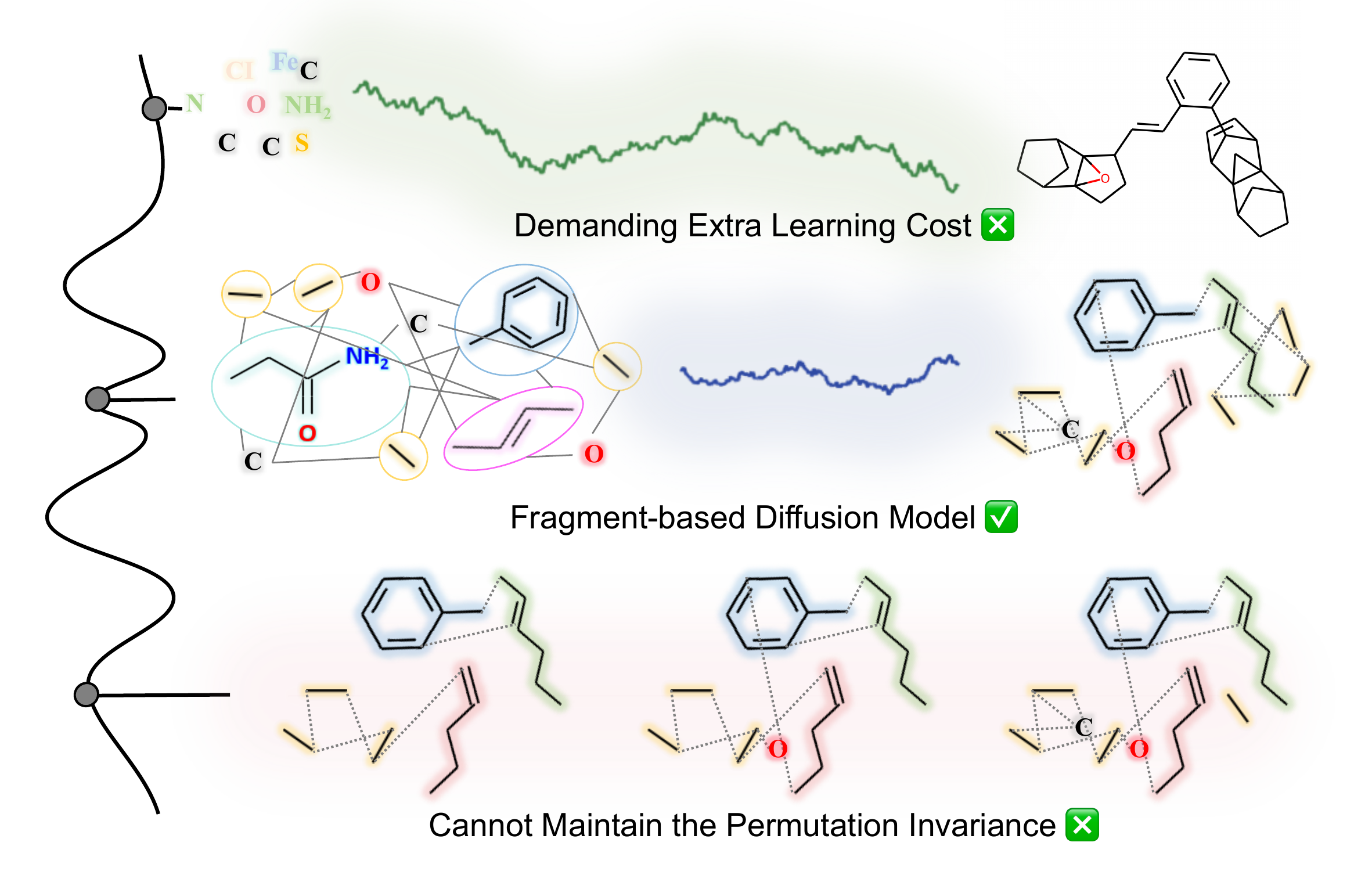}
    \caption{The motivation of our fragment-based diffusion framework. While the atom-level diffusion model demands extra learning cost and the structure-by-structure model cannot maintain the permutation invariance, the fragment-based diffusion framework conquers these challenges.}
    \label{1}
    \vspace{-0.1cm}
\end{figure}

In this paper, we propose OrgMol-Design, a multi-granularity score-based generative framework for the inverse design of complex organics via fragment-level descriptors to easily generate complex organic functional groups and efficiently reduce the large scale of graph nodes compared with atom-level tokens. However, introducing fragment-based descriptions brings additional challenges: Most existing structure-by-structure generation methods require the explicit definition of specific molecular fragment order rules, which disrupts the inherent permutation invariance among fragments. Furthermore, using fragments shifts the complexity of the atomic combinatorial space into the feature space of fragments. Directly applying VAE for one-step generation struggles to capture the manifold within the fragment feature space. Secondly, relying solely on the description of molecular fragments often results in the loss of information about the fragment structures themselves. This hampers the generation of complete molecules and impedes the model's ability to learn the relationship between fragment structures and the global properties of the molecule. In organic molecules, certain crucial functional groups often determine the properties of the molecule. To address the first constraint, we propose a coarse-grained fragment-based diffusion method to achieve a fragment-permutation-invariant generation module. To address the second limitation, we design a fine-grained bond scoring network based on chemical rules. This network facilitates the connections between fragments while also preserving the structural information and molecular properties of the fragments.  We conducted extensive experiments on four real-world benchmarks across different scientific domains, demonstrating OrgMol-Design’s superior performance and efficiency in generating complex organic molecules.

\section{Related Work}

\subsection{Molecular Inverse Design}

Current molecular inverse design models primarily rely on 1D, 2D, or 3D representations. The 1D string-based models typically use the Simplified Molecular Input Line Entry System (SMILES) to describe molecules, such as SMILES-LSTM-HC \cite{doi:10.1021/acs.jcim.8b00839} and GVAE \cite{GVAE}, which model molecules as linear sequences, enabling the application of sequence modeling techniques to inverse design. However, these methods often struggle to capture the complex structural and stereochemical details intrinsic to molecular architectures, leading to challenges in ensuring chemical validity. On the other hand, 3D-based methods ~\cite{hoogeboom2022equivariant, xu2024geometric, qu2024molcraft} focus on the spatial configurations of atoms and provide detailed geometric insights for tasks such as molecular docking. Despite these advantages, these 3D methods generally entail substantial computational overhead and complexity. In contrast, 2D graph-based models present a balanced alternative by representing molecules as graphs, aligning closely with the natural structure of molecules. This representation facilitates the incorporation of topological and chemical features, enabling efficient and accurate molecular generation while mitigating the limitations observed in 1D and 3D approaches.

\subsection{Fragment-based Molecule Design}

Fragment-based molecule design has been explored in previous studies and broadly classified into two categories: chemically inspired and data-driven approaches. Chemically inspired methods rely on hand-crafted rules or external chemical fragment libraries for molecular decomposition. For example, JT-VAE \cite{jtvae} generates molecules as junction trees, with each node representing a ring or an edge. HierVAE \cite{HierVAE} decomposes molecules into subgraphs by severing bridge bonds. FREED \cite{YangHLRH21} extracts fragments from existing chemical fragment libraries. In contrast to these approaches, our approach autonomously extracts frequent fragments to form a vocabulary for segmenting molecules. MiCaM \cite{geng2023de} also attempts fragment mining; however, its vocabulary includes connection information, leading to a substantial increase in the dimensionality of node features, which presents significant computational challenges for diffusion models.

\section{Preliminary}

\subsection{Molecular Fragment Definition}


A molecule is defined as $\mathcal{G} = \{\mathcal{V}, \mathcal{E}\}$, where $\mathcal{V}$ is a set of nodes corresponding to atoms and $\mathcal{E}$ is a set of edges corresponding to chemical bonds. We define a fragment of $\mathcal{G}$ as $\mathcal{F} = \{\tilde{\mathcal{V}}, \tilde{\mathcal{E}}\}$, where $\tilde{\mathcal{V}} \subseteq \mathcal{V}$ and $ \tilde{\mathcal{E}} \subseteq \mathcal{E}$. For fragments $\mathcal{F}$ and $\mathcal{F'}$ from the same molecule, $\mathcal{F} \bigcup \mathcal{F'}$ is the union of fragments $\mathcal{F}$ and $\mathcal{F'}$, together with all edges connecting these two. If an atom in a molecule belongs to both $\mathcal{F}$ and $\mathcal{F'}$, then $\mathcal{F} \cap \mathcal{F'} \neq \emptyset$. Each molecule can be decomposed into a set of fragments $\{\mathcal{F}_i\}_i^k$ and their connections $\{\mathcal{E}_{ij}\}_{i,j}^{k,k}$ when $\mathcal{G} = (\bigcup_i^k\mathcal{F}_i)\bigcup(\bigcup_{i,j}^{k,k}\mathcal{E}_{ij})$ and $\mathcal{F}_i \cap \mathcal{F}_j = \emptyset$ for any $i \neq j$.

\subsection{Score-based Generation Formulation}


We define a fragment-level molecule graph $\textbf{\em G}^{\mathcal{F}}=(\textbf{\em F}, \textbf{\em C})$ with fragment features $\textbf{\em F} \in \mathbb{R}^{N \times K}$ and adjacency matrix representing fragment connections $\textbf{\em C} \in \mathbb{R}^{N \times N}$, where $N$ is the number of fragments, and $K$ is the vocabulary size. The graph generation process is modeled over time steps $T$, starting with $\textbf{\em G}^{\mathcal{F}}_0$ sampled from the dataset distribution $p_{dataset}$. A noisy trajectory is defined as $\{\textbf{\em G}^{\mathcal{F}}_t=(\textbf{\em F}_t, \textbf{\em C}_t)\}_{t \in [0,T]}$, where $[0, T]$ is the time step range. This process aims to generate complex organic functional groups while reducing the graph size relative to atom-level tokens. The resulting fragments and their connections are then refined in the bond-scoring module.

\section{Methodology}

In this section, we present the details of our proposed OrgMol-Design. We first elaborate on the process of constructing the fragment vocabulary. Then, we explain the multi-granularity framework for inverse design of complex organics, consisting of a score-based generative model via fragment prior to achieve diverse coarse-grained scaffold generation, and a bond scoring model guided by valence and cycle rules for fine-grained molecular structure design.

\subsection{Fragment Vocabulary Construction}

We aim to identify frequent molecular fragments in the training dataset to build a comprehensive fragment vocabulary for molecule inverse design. Inspired by Byte Pair Encoding \cite{kong2022molecule}, we employ a bottom-up mechanism that iteratively merges frequent adjacent fragment pairs, starting from single atoms, to generate more complex fragments. Appendix A provides an illustrative example and a detailed pseudo code. Atom proportions in fragments across vocabulary sizes are in Appendix E.

The vocabulary is initialized with individual atoms (denoted as $\mathcal{N}_{atom}$ for the total number of atoms). Given a target vocabulary size $\mathcal{N}$, we conduct $\mathcal{N}-\mathcal{N}_{atom}$ iterations to construct the complete vocabulary. In each iteration, we merge every neighboring fragments $\mathcal{F}$ and $\mathcal{F'}$ by forming the union $\mathcal{F} \bigcup \mathcal{F'}$. For a given fragment $\mathcal{F} = \{\tilde{\mathcal{V}}, \tilde{\mathcal{E}}\}$ in a molecule, its neighbors can be represented as $\mathcal{S}_{\mathcal{F}} = \{\mathcal{F'}:= \{\tilde{\mathcal{V'}}, \tilde{\mathcal{E'}}\} \! \mid \! \exists \, v \in \tilde{\mathcal{V}}, v' \in \tilde{\mathcal{V'}}, d(v, v') = 1 \}$, where $d(v, v')$ is the shortest distance in graph topology space between nodes $v'$ and $v$. Then, we record the frequency of each merged fragment and add the most frequent one to the vocabulary. Finally, we repeat the previous steps from merging until the vocabulary reaches size $\mathcal{N}$.

\subsection{Multi-granularity Molecule Inverse Design}

\begin{figure*}[t]
    \centering
    \includegraphics[width=1\linewidth]{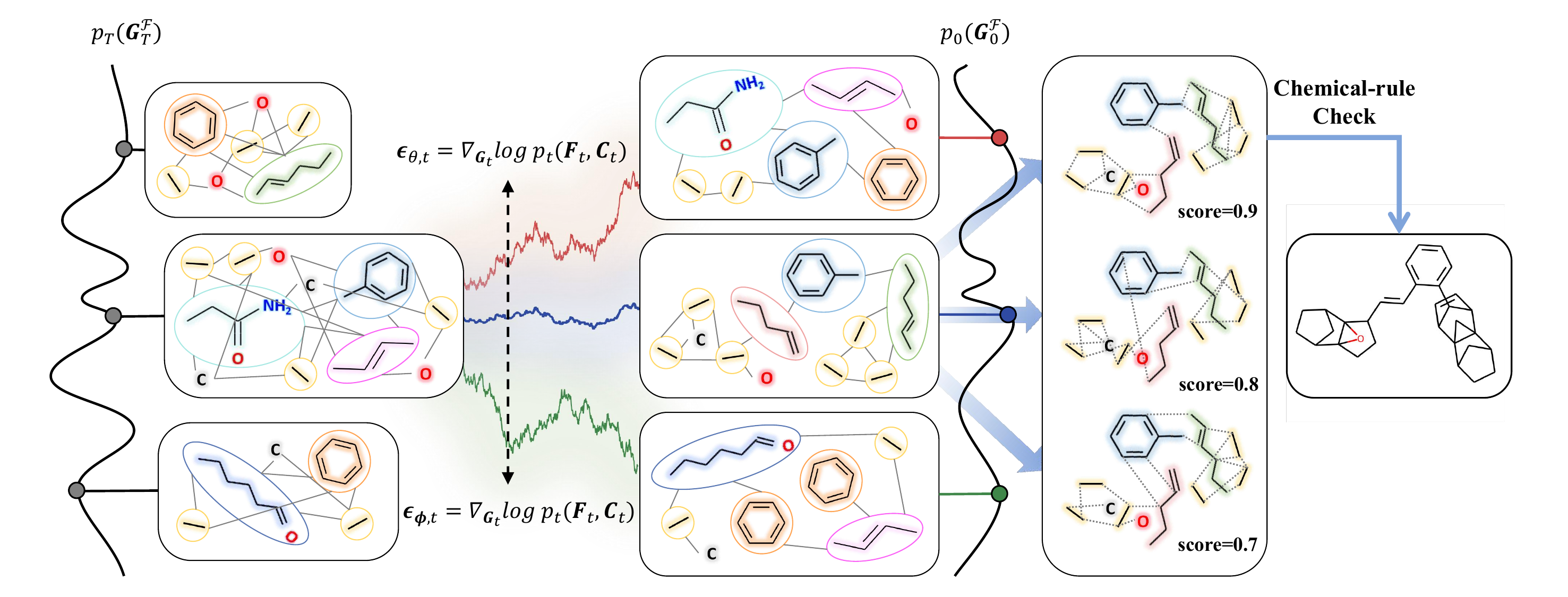}
    \caption{An overview of OrgMol-Design. (Left) Coarse-grained fragment generation. Sampling randomly connected fragments from the prior distribution at $t=T$. Colored trajectories represent different diffusion processes in the joint space of fragment features and connections. (Middle) Generated fragments and connections at $t=0$. (Right) Fine-grained bond scoring. The highest-scoring connection is selected, completing the molecule after a chemical-rule check.}
    \label{model}
    \vspace{-0.5cm}
\end{figure*}

We formulate the molecule inverse design as a multi-granularity process: (1) generating a coarse-grained molecular scaffold based on fragment tokens and their connections, and (2) scoring the fine-grained connection sites for these connected fragments to assemble a complete molecule. An overview is provided in Figure \ref{model}. 

\subsubsection{Coarse-grained Score-based Generative Modeling}

The goal is to generate fragment graphs mirroring the distribution of observed decomposed molecules. We introduce a continuous-time graph diffusion process that transforms fragment features and adjacency matrices into a known prior distribution and back, capturing fragment-connection dependencies.

Formally, given a trajectory of noised graph variables $\{\textbf{\em G}^{\mathcal{F}}_t=(\textbf{\em F}_t, \textbf{\em C}_t)\}_{t \in (0,T]}$, where $\textbf{\em G}^{\mathcal{F}}_0 \sim p_{dataset}$, our forward process follows stochastic differential equations (SDEs):
\begin{equation}
    \mathrm{d}\textbf{\em G}^{\mathcal{F}}_t = \textbf{f}_t(\textbf{\em G}^{\mathcal{F}}_t)\mathrm{d}t+\textbf{g}_t(\textbf{\em G}^{\mathcal{F}}_t)\mathrm{d}\textbf{w}, \label{forward}
\end{equation}
where $\textbf{f}_t(\cdot)$ and $\textbf{g}_t(\cdot)$ are the linear drift and diffusion coefficients, and $\textbf{w}$ represents the standard Wiener process \cite{karlin1981stochastic}. For simplicity, we choose $\textbf{g}_t(\textbf{\em G}^{\mathcal{F}}_t)$ as a scalar function $g_t$. By adding noise $\mathrm{d}\textbf{w}$ at each time step $\mathrm{d}t$, $\textbf{\em F}_0$ and $\textbf{\em C}_0$ are jointly transformed to a prior distribution (e.g., Gaussian). To generate fragment graphs, we sample from this prior and reverse the diffusion process: 
\begin{equation}
    \mathrm{d}\textbf{\em G}^{\mathcal{F}}_t = [\textbf{f}_t(\textbf{\em G}^{\mathcal{F}}_t)-g_t^2\nabla_{\textbf{\em G}^{\mathcal{F}}_t}\mathrm{log}\,p_t(\textbf{\em G}^{\mathcal{F}}_t)]\mathrm{d}\bar{t}+g_t\mathrm{d}\bar{\textbf{w}}, \label{reverse}
\end{equation}
where $\mathrm{d}\bar{t}$ is the negative time steps from $T$ to $0$, $\bar{\textbf{w}}$ is the reverse Wiener process, and $p_t$ is the marginal distribution. To reduce computational load, we decompose Equation \ref{reverse} into the node and topology components:
\begin{equation}
\begin{aligned}
    & \mathrm{d}\textbf{\em F}_t = [\textbf{f}_{1,t}(\textbf{\em F}_t)-g_{1,t}^2\nabla_{\textbf{\em F}_t}\mathrm{log}\,p_t(\textbf{\em G}^{\mathcal{F}}_t)]\mathrm{d}\bar{t}+g_{1,t}\mathrm{d}\bar{\textbf{w}}_1, \\
    & \mathrm{d}\textbf{\em C}_t = [\textbf{f}_{2,t}(\textbf{\em C}_t)-g_{2,t}^2\nabla_{\textbf{\em C}_t}\mathrm{log}\,p_t(\textbf{\em G}^{\mathcal{F}}_t)]\mathrm{d}\bar{t}+g_{2,t}\mathrm{d}\bar{\textbf{w}}_2, 
    \label{decompose}
\end{aligned}
\end{equation}
where $\textbf{f}_{1,t}$ and $\textbf{f}_{2,t}$ are linear drift coefficients that satisfy $\textbf{f}_t(\textbf{\em F}, \textbf{\em C}) = (\textbf{f}_{1,t}(\textbf{\em F}), \textbf{f}_{2,t}(\textbf{\em C}))$. Similarly, $g_{1,t}$ and $g_{2,t}$ are scalar diffusion coefficients, while $\bar{\textbf{w}}_1$ and $\bar{\textbf{w}}_2$ denote the reverse-time standard Wiener processes. 
We train two score matching neural networks, $\boldsymbol{\epsilon}_{\theta,t}$ and $\boldsymbol{\epsilon}_{\phi,t}$, to parameterize the component-wise scores in Equation \ref{decompose}, where the former estimates the node component $\nabla_{\textbf{\em F}_t}\mathrm{log}\,p_t(\textbf{\em F}_t, \textbf{\em C}_t)$, while the latter estimates the topology component $\nabla_{\textbf{\em C}_t}\mathrm{log}\,p_t(\textbf{\em F}_t, \textbf{\em C}_t)$. The model minimizes the distance to the ground-truth component-wise scores through the following training objectives:
\begin{equation}
\begin{aligned}
    & \min \limits_{\theta} \mathbb{E}_t \{ \tau_1(t)\mathbb{E}_{\textbf{\em G}^{\mathcal{F}}_0}\mathbb{E}_{\textbf{\em G}^{\mathcal{F}}_t \mid \textbf{\em G}^{\mathcal{F}}_0} \parallel \boldsymbol{\epsilon}_{\theta,t} (\textbf{\em G}^{\mathcal{F}}_t) - \nabla_{\textbf{\em F}_t}\mathrm{log}\,p_t(\textbf{\em G}^{\mathcal{F}}_t) \parallel_2^2 \}, \\
    & \min \limits_{\phi} \mathbb{E}_t \{ \tau_2(t)\mathbb{E}_{\textbf{\em G}^{\mathcal{F}}_0}\mathbb{E}_{\textbf{\em G}^{\mathcal{F}}_t \mid \textbf{\em G}^{\mathcal{F}}_0} \parallel \boldsymbol{\epsilon}_{\phi,t} (\textbf{\em G}^{\mathcal{F}}_t) - \nabla_{\textbf{\em C}_t}\mathrm{log}\,p_t(\textbf{\em G}^{\mathcal{F}}_t) \parallel_2^2 \}, 
    \label{objectives}
\end{aligned}
\end{equation}
where $\tau_1(t)$ and $\tau_2(t)$ are weighting functions. However, Equation \ref{objectives} is not directly applicable for training, as the ground-truth scores are analytically insoluble. Inspired by \cite{song2021scorebased}, we substitute $p_t(\textbf{\em G}^{\mathcal{F}}_t)$ with $p_{0t}(\textbf{\em G}^{\mathcal{F}}_t \mid \textbf{\em G}^{\mathcal{F}}_0)$, where $\textbf{\em G}^{\mathcal{F}}_0 \sim p_{dataset}$ and $\textbf{\em G}^{\mathcal{F}}_t \sim p_{0t}(\textbf{\em G}^{\mathcal{F}}_t \mid \textbf{\em G}^{\mathcal{F}}_0)$. The transition distribution $p_{0t}(\textbf{\em G}^{\mathcal{F}}_t \mid \textbf{\em G}^{\mathcal{F}}_0)$, driven by the forward diffusion process, decomposes as follows: 
\begin{equation}
    p_{0t}(\textbf{\em G}^{\mathcal{F}}_t \mid \textbf{\em G}^{\mathcal{F}}_0) = p_{0t}(\textbf{\em F}_t \mid \textbf{\em F}_0) \, p_{0t}(\textbf{\em C}_t \mid \textbf{\em C}_0). \label{linearity}
\end{equation}

Sampling from $p_{0t}(\textbf{\em F}_t \mid \textbf{\em F}_0)$ and $p_{0t}(\textbf{\em C}_t \mid \textbf{\em C}_0)$ is effortless, as they are Gaussian distributions with mean and variance defined by the forward diffusion process coefficients. The corresponding training objectives are then derived as:
\begin{equation}
\begin{aligned}
    & \min \limits_{\theta} \mathbb{E}_t \{ \tau_1(t)\mathbb{E}_{\textbf{\em G}^{\mathcal{F}}_0}\mathbb{E}_{\textbf{\em G}^{\mathcal{F}}_t \mid \textbf{\em G}^{\mathcal{F}}_0} \! \parallel \!\boldsymbol{\epsilon}_{\theta,t} (\textbf{\em G}^{\mathcal{F}}_t) - \nabla_{\textbf{\em F}_t}\mathrm{log}p_{0t}(\textbf{\em F}_t \mid \textbf{\em F}_0) \parallel_2^2 \}, \\
    & \min \limits_{\phi} \mathbb{E}_t \{ \tau_2(t)\mathbb{E}_{\textbf{\em G}^{\mathcal{F}}_0}\mathbb{E}_{\textbf{\em G}^{\mathcal{F}}_t \mid \textbf{\em G}^{\mathcal{F}}_0} \! \parallel \!\boldsymbol{\epsilon}_{\phi,t} (\textbf{\em G}^{\mathcal{F}}_t) - \nabla_{\textbf{\em C}_t}\mathrm{log}p_{0t}(\textbf{\em C}_t \mid \textbf{\em C}_0) \parallel_2^2 \}. 
\end{aligned}
\label{new_objectives}
\end{equation}

By minimizing the aforementioned objectives, we can effectively estimate the component-wise scores. To address simultaneously model $\textbf{\em F}_t$ and $\textbf{\em C}_t$ along with their interdependencies, we propose various graph neural network (GNN) architectures tailored to this task.

First, for the model $\boldsymbol{\epsilon}_{\theta,t}$ that estimates $\nabla_{\textbf{\em F}_t}\mathrm{log}\,p_t(\textbf{\em F}_t, \textbf{\em C}_t)$, we utilize multiple layers of graph convolutional networks (GCNs) \cite{kipf2017semisupervised} structured as follows:
\begin{equation}
\begin{aligned}
    \boldsymbol{\epsilon}_{\theta,t}(\textbf{\em G}^{\mathcal{F}}_t&) = \mathrm{MLP} ([\{\textbf{\em H}_i\}_{i=1}^L]), \\
    \textbf{\em H}_{i} & = \mathrm{GCN}(\textbf{\em H}_{i-1}, \textbf{\em C}_t),
\end{aligned}
\label{scoreX}
\end{equation}
where $\textbf{\em H}_0 = \textbf{\em F}_t$ and $L$ is the number of GCN layers. 

For the model $\boldsymbol{\epsilon}_{\phi,t}$ which estimates $\nabla_{\textbf{\em C}_t}\mathrm{log}\,p_t(\textbf{\em F}_t, \textbf{\em C}_t)$, we leverage graph multi-head attention (GMH) \cite{DBLP:journals/corr/abs-2102-11533} with high-order adjacency matrices:
\begin{equation}
\begin{aligned}
    \boldsymbol{\epsilon}_{\phi,t}(\textbf{\em G}^{\mathcal{F}}_t) = \mathrm{MLP} \big(  &\big[ \{ \mathrm{GMH} (\textbf{\em H}_i , \textbf{\em C}_t^d)\}_{i=1,d=1}^{M,D} \big] \big), \\
    \textbf{\em H}_{i} = & \; \mathrm{GCN}(\textbf{\em H}_{i-1}, \textbf{\em C}_t),
    \label{scoreA}
\end{aligned}
\end{equation}
where $\textbf{\em H}_0 = \textbf{\em F}_t$, and $\textbf{\em C}_t^d$ is the high-order adjacency matrices with a total order $D$. $M$ is the number of GMH layers. GMH models the fragment interactions based on topology, and high-order adjacency matrices capture far-reaching dependencies.

To generate fragment graphs from the parameterized SDEs in Equation \ref{decompose}, we use numerical solvers, specifically the Predictor-Corrector Sampler (PC Sampler) \cite{song2021scorebased}, which efficiently explores high-density data regions. During the prediction stage, we utilize SDEs such as Variance Exploding (VE) and Variance Preserving (VP) \cite{song2021scorebased}. For the correction phase, we implement the Langevin Markov Chain Monte Carlo (MCMC) \cite{PARISI1981378}. For more details on coarse-grained score-based generative modeling, refer to Appendix B.

\subsubsection{Fine-grained Bond Scoring Model}

This stage aims to assemble the fragments at a finer granularity, involving bond connections between fragment pairs. To accomplish this, we propose a bond scoring model based on GNNs that non-autoregressively and globally estimates possible chemical bond types between fragment pairs following chemical rules. Each bond type is assigned a score, enabling the selection of the most appropriate predicted edges to construct the final complex organic molecules.

Specifically, given nodes $u \in \mathcal{F}_i$ and $v \in \mathcal{F}_j$ from different fragments in a molecule graph, their edge connection is estimated as:
\begin{equation}
\begin{aligned}
    \textbf{\em p}(e^{uv}) & = \boldsymbol{\delta}_{\vartheta}  ([\textbf{\em H}^u_k \parallel \textbf{\em H}^v_k]), \\
    \textbf{\em H}_{i} = \mathrm{GINE} & ( \mathrm{MLP} (\textbf{\em H}_{i-1}, \textbf{\em C}_{i-1})), \\
    \textbf{\em C}_{i} = & \; \mathrm{Linear} (\textbf{\em C}_{i-1}),
\label{uv}
\end{aligned}
\end{equation}
where $i \in [1, k]$, $\textbf{\em H}_0 = \textbf{\em F} \in \mathbb{R}^{N \times d}$, and $\textbf{\em C}_0 = \textbf{\em C} \in \mathbb{R}^{N \times N \times d'}$ with $N$ denoting the number of atoms in the molecule. Additionally, $d$ and $d'$ are the dimensions of atomic nodes and edges, respectively, and $\boldsymbol{\delta}_{\vartheta}$ is a two-layer MLP with ReLU activation. We employ GINE \cite{hu2020pretraining} for message-passing to obtain node embeddings.

Considering the undirected nature of chemical bonds, we compute both $\textbf{\em p}(e^{uv})$ and $\textbf{\em p}(e^{vu})$. Besides the three standard chemical bonds (i.e., single, double, and triple bonds), we introduce a "none" type to indicate the absence of a bond. Due to the predominance of "none" bonds, we employ negative sampling \cite{word2vec} to mitigate the information loss caused by this imbalance. The loss function for this stage is defined as follows: 
\begin{equation}
    \mathcal{L} = \sum \limits_{u \in \mathcal{F}_i , v \in \mathcal{F}_j , i \neq j} - \, \mathrm{log} \, \textbf{\em p}(e^{uv}) . \label{loss}
\end{equation}

To generate complete molecules in the inference phase, we devise a scoring mechanism grounded in chemical rules to decode the predicted edges. From the coarse-grained fragment generation module, we obtain a fragment set $\{\mathcal{F}_i\}_i^m$ and a fragment-level adjacency matrix $\textbf{\em C} \in \mathbb{R}^{m \times m}$. If a connection exists between fragments $\mathcal{F}_i$ and $\mathcal{F}_j$, then $\textbf{\em C}_{ij} = \textbf{\em C}_{ji} = 1$; otherwise, $\textbf{\em C}_{ij} = \textbf{\em C}_{ji} = 0$. For each independent molecular fragment, we initially utilize RDKit \cite{Landrum2016RDKit2016_09_4} to gradually add its atoms and chemical bonds to the partially constructed molecular structure, while concurrently recording the intra-fragment edge sets $\{\mathcal{E}_i^{intra}\}_i^m$. This approach yields an incomplete molecule without inter-fragment connections. 

Assuming each atom is mapped to its corresponding fragment via $\boldsymbol{\omega}$, we define candidate inter-fragment edges as:
\begin{equation}
\begin{aligned}
    \textbf{\em E}(u, v) = 
    \begin{cases}
    0 \quad \mathrm{if} \; \exists \, \mathcal{E}_k^{intra} \, \mathrm{such \, that} \, (u,v) \in \mathcal{E}_k^{intra}, \\
    0 \quad \mathrm{if} \; \textbf{\em C}(\boldsymbol{\omega}(u),\boldsymbol{\omega}(v))=0,    \\
    1 \quad \mathrm{otherwise},
    \end{cases}
\end{aligned}
\label{bond}
\end{equation}
where $\textbf{\em E}(u, v) = 1$ indicates a potential bond between nodes $u$ and $v$ in different fragment. We compute scores $\mathcal{J}$ for candidate edges as:
\begin{equation}
\begin{aligned}
    \mathcal{J}(e^{uv}) = \mathrm{max} & ( \mathrm{softmax} (\textbf{\em p}(e^{uv}))), \\
    \mathrm{softmax} (\textbf{\em p}(e^{uv}_k)) & = \frac{\mathrm{exp}(\textbf{\em p}(e^{uv}_k))}{\sum_{l=1}^{E}\mathrm{exp}(\textbf{\em p}(e^{uv}_l)},
\end{aligned}
\label{score}
\end{equation}
where $\textbf{\em p}(e^{uv}) \in \mathbb{R}^{1 \times E}$ is the prediction of edge types, and $E$ is the total number of edge types. Subsequently, we apply a softmax function to transform the prediction into confidence values, reflecting the probability that $e^{uv}$ belongs to each possible type. The score for $e^{uv}$ is then defined as the maximum among these confidence values.

\begin{table}[t]
    \centering
    \fontsize{9}{9}\selectfont 
    \begin{tabular}{@{}lc|ccc@{}}
    \toprule
    \textbf{Methods} & \shortstack{\textbf{PCE\textsubscript{PCBM}} \\ \textbf{-SAscore}} & \shortstack{\textbf{PCE\textsubscript{PCDTBT}} \\ \textbf{-SAscore}} \\ 
    \midrule
    Dataset & 7.57 & 31.71  \\
    SMILES-VAE & 7.44±0.28  & 10.23±11.14  \\
    SMILES-LSTM-HC & 6.69±0.40  & \textbf{31.79±0.15}  \\
    MoFlow  & 7.08±0.31  & 29.81±0.37  \\
    REINVENT & 7.48±0.11  & 30.47±0.44  \\
    GB-GA & 7.78±0.02  & 30.24±0.80  \\
    GDSS & 1.37±0.34  & 25.05±1.05  \\
    MiCaM & 3.96±0.37  & 23.99±0.91  \\
    \midrule
    \textbf{OrgMol-Design} & \textbf{7.98±0.16} & 30.01±0.37 \\
    \bottomrule
    \end{tabular}
    \caption{Results for the organic photovoltaics design benchmark, mean±std of the best objective values obtained from five independent runs.}
    \label{table_hce}
    \vspace{-0.5cm}
\end{table}

We rank the candidate edges in descending order according to their scores. Then, we iterate through these edges, adding an edge to the molecule if its score exceeds the threshold $\Psi_{th}$, provided it satisfies the chemical-rule check. The chemical check ensures that the proposed bond adheres to valence rules, and does not form an unstable ring composed of fewer than five or more than six nodes. Due to the possibility of generating disconnected graphs during this process, we identify the largest connected component as the final organic molecule. The pseudo code of the above fine-grained algorithm is in Appendix C. 

\begin{table*}[t]

    \centering
    \fontsize{9}{9}\selectfont
    \begin{tabular}{@{}lc|c|c|c@{}}
    \toprule
    \textbf{Methods} & $\boldsymbol{\Delta E_{a}}$ & $\boldsymbol{\Delta E_{r}}$ & $\boldsymbol{\Delta E_{a}+\Delta E_{r}}$ & $\boldsymbol{-\Delta E_{a}+\Delta E_{r}}$\\ 
    \midrule
    Parent Substrate & 85.16 & 0.00 & 85.16 & -85.16  \\
    Dataset & 64.94 & -34.39 & 56.48 & -95.25  \\
    SMILES-VAE & 76.81±0.25  & -10.96±0.71 & 71.01±0.62 & -90.94±1.04  \\
    SMILES-LSTM-HC & 59.64±4.10  & -31.03±16.15 & 71.81±1.56 & -91.58±2.14  \\
    MoFlow  & 70.12±2.13  & -20.21±4.13 & 63.21±0.69 & -92.82±3.06  \\
    REINVENT & 68.38±2.00  & -24.35±6.46 & 55.25±5.88 & -94.52±1.20  \\
    GB-GA & 56.04±3.07  & -41.39±5.76 & 45.20±6.78 & -100.07±1.35  \\
    GDSS & -  & - & - & -  \\
    MiCaM & 73.92±1.34  & -15.48±1.54 & 60.16±2.80 & -95.40±1.03  \\
    \midrule
    \textbf{OrgMol-Design} & \textbf{44.47±2.31} & \textbf{-49.36±0.26} & \textbf{15.75±2.50} & \textbf{-108.28±3.35} \\
    \bottomrule
    \end{tabular}
    \caption{Results for the molecular reactivity benchmark, mean±std of optimal objective values over five independent runs.}
    \label{table_snb}

\end{table*}

\section{Experiments}

In this section, we showcase the performance of OrgMol-Design across various benchmarks in real-world scenarios, including complex organics design in optoelectronic materials, catalyst for organic reaction optimization, and ligand for protein target. Additionally, we also demonstrate the efficiency of our model compared with previous diffusion-based models. In Appendix D, we provide detailed results of ablation studies which show how different modules, such as fine-grained bond scoring, fragment-level connections, etc., affect the quality of the generated molecules. 

\subsection{Experimental Setup}

To rigorously assess the performance of OrgMol-Design, we have selected four benchmark datasets (HCE, GDB-13, SNB-60K, and DTP, as detailed in Appendix I.1) where molecules either contain complex functional groups or possess a large number of atoms. In addition to standard metrics (detailed in Appendix F) such as novelty, validity, and uniqueness, our primary focus is on the ideal combinations of various properties of the generated molecules. Hyperparameter settings are provided in Appendix H.

\begin{table}[t]
    \centering
    \fontsize{9}{9}\selectfont
    \begin{tabular}{@{}lc|c|cc@{}}
    \toprule
    \textbf{Methods} & \textbf{ST} & \textbf{OSC} & \textbf{Combined}\\ 
    \midrule
    Dataset & 0.020 & 2.97 & -0.04  \\
    SMILES-VAE & 0.071±0.003  & 0.50±0.27 & -0.57±0.33  \\
    SMILES-LSTM-HC & 0.015±0.002  & 1.00±0.01 & -0.24±0.01  \\
    MoFlow  & 0.013±0.001  & 0.81±0.11 & -0.04±0.06  \\
    REINVENT & 0.014±0.003  & 1.16±0.18 & -0.15±0.05  \\
    GB-GA & 0.012±0.002  & 2.14±0.45 & 0.07±0.03  \\
    GDSS & 0.008±0.007  & 1.40±0.11 & -0.27±0.03  \\
    MiCaM & 0.006±0.004  & 1.26±0.17 & -0.12±0.06  \\
    \midrule
    \textbf{OrgMol-Design} & \textbf{0.002±0.001} & \textbf{2.40±0.30} & \textbf{0.61±0.49} \\
    \bottomrule
    \end{tabular}
    \caption{Results for the organic emitters design benchmark, provided as mean±std of the best target objective values from five independent runs.}
    \label{table_gdb13}
\end{table}

\begin{table}[t]
    \centering
    \fontsize{9}{9}\selectfont 
    \begin{tabular}{@{}lc|c|cc@{}}
    \toprule
    \textbf{Methods} & \textbf{DS\textsubscript{qvina}} & \textbf{DS\textsubscript{smina}} & \textbf{SR}\\ 
    \midrule
    Native Docking & -11.6 & -12.1 & 100.0\%  \\
    Dataset & -12.2 & -13.1 & 100.0\%  \\
    SMILES-VAE & -10.7±0.2  & -11.1±0.4 & 12.6\%  \\
    SMILES-LSTM-HC & -12.4±0.3  & -13.3±0.4 & 73.9\%  \\
    MoFlow  & -12.1±0.4  & -13.0±0.3 & 36.2\%  \\
    REINVENT & -12.8±0.2  & -13.7±0.5 & 76.8\%  \\
    GB-GA & -12.9±0.1  & -13.8±0.4 & 71.4\%  \\
    GDSS & -10.56±0.6  & -11.04±0.2 & 98.9\%  \\
    MiCaM & -11.54±0.4  & -11.66±0.3 & 99.1\%  \\
    \midrule
    \textbf{OrgMol-Design} & \textbf{-13.48±0.2} & \textbf{-14.26±0.9} & \textbf{99.4\%} \\
    \bottomrule
    \end{tabular}
    \caption{Results for the protein ligands design benchmark, where the docking scores are mean±std of the best values from five independent runs.}
    \label{table_dtp}
    \vspace{-0.5cm}
\end{table}

We compared OrgMol-Design against a diverse set of molecular generative models. SMILES-VAE \cite{doi:10.1021/acscentsci.7b00572} employs a VAE framework based on the SMILES representation. SMILES-LSTM-HC \cite{doi:10.1021/acs.jcim.8b00839} is a long short-term memory-based model that utilizes hill-climbing within the SMILES framework. MoFlow \cite{10.1145/3394486.3403104} is a flow-based generative model. REINVENT \cite{Olivecrona2017MolecularDD} is a reinforcement learning-based model that leverages a recurrent neural network. GB-GA \cite{C8SC05372C} is a graph-based genetic algorithm that simulates atomic and bond distributions during genetic operations. GDSS \cite{jo2022GDSS} is a diffusion-based model with atoms as the small descriptive units, which generates nodes and edges interdependently. MiCaM \cite{geng2023de} is a fragment-based VAE model. The settings for the above baseline methods are provided in Appendix I.6.

\subsection{Design of Organic Photovoltaics}

Organic photovoltaics (OPVs) are pivotal in advancing renewable energy through improved efficiency, cost-effectiveness, and versatility in organic solar cells (OSCs). Despite progress, challenges persist with low power conversion efficiencies (PCEs). To tackle this, we introduced two benchmark tasks trained on the HCE dataset to discover novel organic photoactive materials with superior PCEs. The first task aims to design a small organic donor molecule paired with PCBM as the acceptor, while the second focuses on designing a small organic acceptor molecule with PCDTBT as the donor. Both tasks evaluate PCEs and synthetic accessibility (SA) scores. The simulation workflow is detailed in Appendix I.2.

Results presented in Table \ref{table_hce} highlight top-performing molecules, including the best molecules from the training set ("Dataset"). OrgMol-Design attains the highest metric value corresponding to PCBM and closely matches the best performance obtained by SMILES-LSTM-HC for PCDTBT. 
SMILES-LSTM-HC's strong results may stem from its ability to model long-range dependencies through string sequences, which can be advantageous for specific tasks. 
Notably, SMILES-VAE and SMILES-LSTM-HC excel in one task but underperform in the other, likely due to the limitations of sequence-based methods in capturing sample features, leading to variability in task performance. In contrast, graph-based models like MoFlow and GB-GA offer better global search capabilities and stability.

\subsection{Design of Chemical Reaction Substrates}

Given the importance of chemical reactions, we employed the benchmark using the SNB-60K dataset to target the modification of substrate and product structures to influence reactivity. We modeled the intramolecular concerted double hydrogen transfer reaction of syn-sesquinorbornenes. In this task, four evaluation metrics are considered: activation energy (related to reaction speed), reaction energy (related to thermodynamic favorability), and their sum and difference. The complete workflow is outlined in Appendix I.3.

Table \ref{table_snb} provides the performance on the molecular reactivity benchmark, where the baselines include the best-performing molecule in the training dataset ("Dataset") and the parent unsubstituted substrate ("Parent Substrate"). The results indicate that our model consistently generates molecules with optimal values on all metrics, especially excelling in the sum of activation and reaction energies. This highlights the model's superiority in designing substrates that promote both rapid and thermodynamically favorable reactions. Notably, the previous diffusion-based model, GDSS, fails to generate valid molecules that pass through structural constraints, underscoring the challenges of atom-level diffusion to learn the potential chemical plausibility compared to our fragment-level approach.

\subsection{Design of Organic Emitters}

Organic light-emitting diodes (OLEDs) have garnered widespread attention since the discovery of thermally activated delayed fluorescence (TADF). To advance OLED design, we employed a benchmark using the GDB-13 dataset to develop organic emissive molecules. Evaluation metrics include the singlet-triplet gap (ST) for efficiency, oscillator strength (OSC) for fluorescence rates, and a combined metric ensuring emission within the blue light energy range. The SA score must not exceed 4.5 for optimal fitness. The entire simulation workflow is detailed in Appendix I.4.

Table \ref{table_gdb13} compares OrgMol-Design's results with the best baseline values. OrgMol-Design achieves state-of-the-art performance across all metrics, demonstrating its ability to generate stable organic emitters with elevated properties. Particularly for the ST and Combined values, our model can successfully generate organics that improve upon the best molecules in the training dataset. The substantial improvement over GDSS further underscores the advantage of using fragments to explore higher-dimensional feature spaces for desired molecule generation.

\subsection{Design of Protein Ligands}

Designing small molecule ligands for specific proteins is of paramount importance in advancing targeted therapy by modulating disease-related proteins. Consequently, we developed a benchmark for ligand design targeting a specific protein (e.g., 4LDE $\beta$2-adrenoceptor GPCR receptor) using molecular docking simulations. Evaluation metrics include docking scores (DS) from QuickVina2 and Smina, which assess binding quality, and the success rate (SR) for molecules passing standard structure filters. Details are in Appendix I.5.

\begin{figure}[t]
    \centering
    \includegraphics[width=1\linewidth]{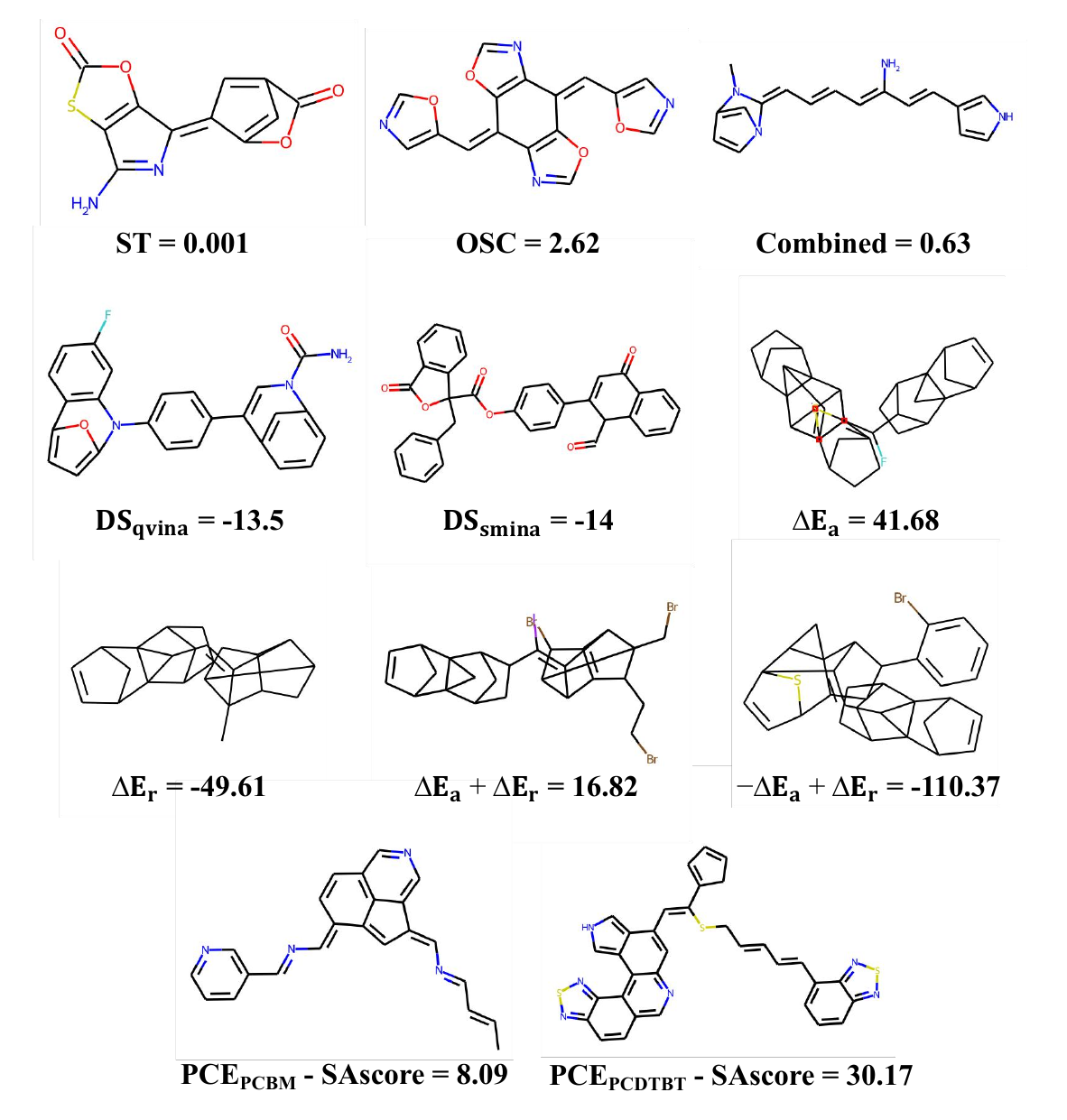}
    \vspace{-0.3cm}
    \caption{Examples of generated molecules corresponding to each metric value.}
    \label{mol}
    \vspace{-0.5cm}
\end{figure}

The performance of the models, trained on the DTP dataset, is summarized in Table \ref{table_dtp}, where the values for "Dataset" and "Native Docking" correspond to the top molecules from the training dataset and the original ligands, respectively. OrgMol-Design outperforms other methods across all metrics, generating ligands with strong binding affinity and a higher SR, indicating stability and synthesizability. Besides, GDSS and MiCaM also achieve SR values over 98\%, indicating the importance of both score-based generation and fragmentation in producing structurally stable and synthesizable molecules.

Figure \ref{mol} illustrates representative molecules generated by OrgMol-Design, corresponding to specific evaluation metrics discussed above.

\begin{figure}[t]
    \centering
    \includegraphics[width=1\linewidth]{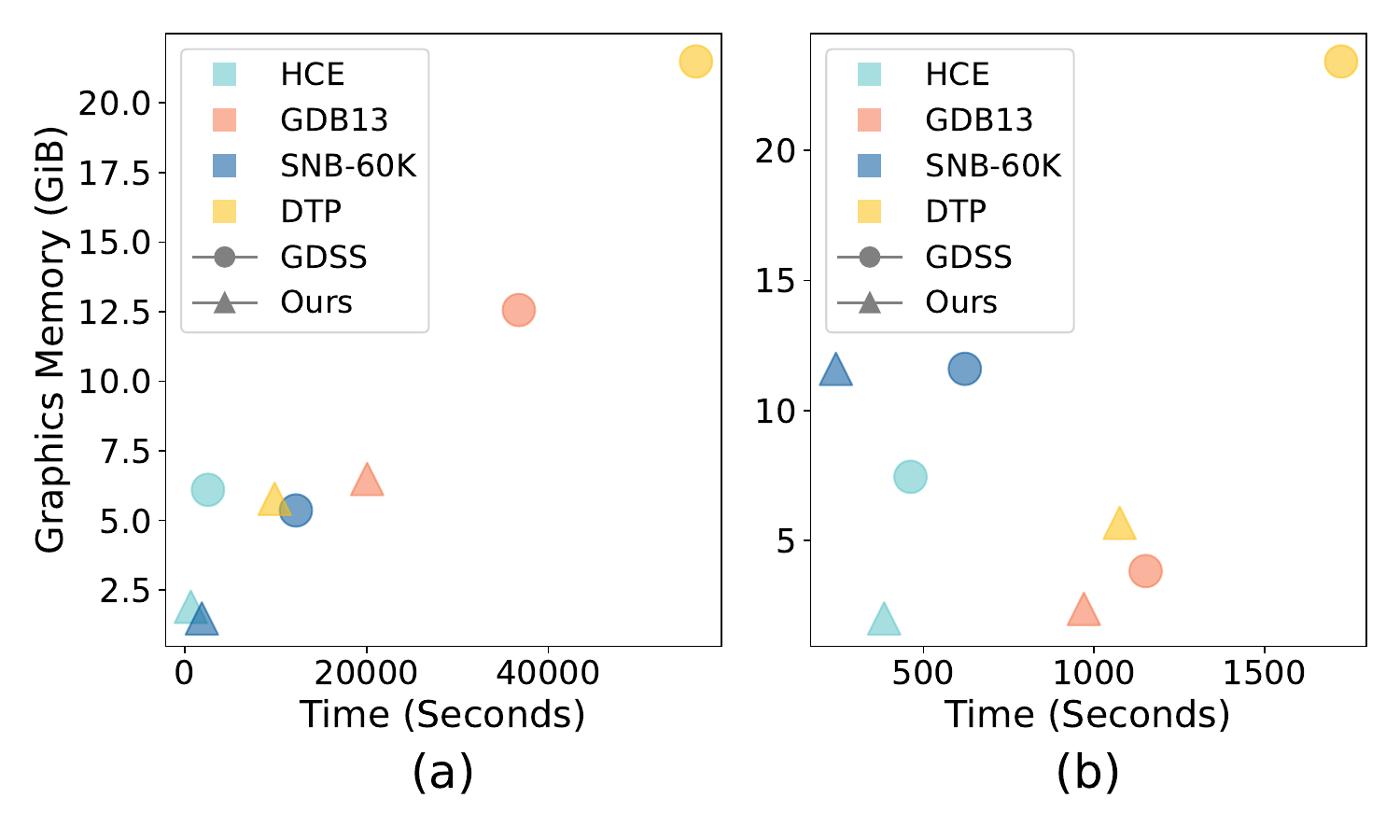}
    \vspace{-0.5cm}
    \caption{Time and graphics memory cost comparison between OrgMol-Design and GDSS during (a) diffusion training and (b) sampling of generating molecules.}
    \label{efficient}
    \vspace{-0.5cm}
\end{figure}

\subsection{Efficiency Analysis}

To verify the efficiency of OrgMol-Design, we conduct comparative experiments against GDSS on an NVIDIA GeForce RTX 3090 GPU, regarding the time and graphics memory usage during diffusion training and molecular sampling. For the HCE, GDB-13, SNB-60K, and DTP datasets, the vocabulary sizes are set to 100, 200, 100, and 200, respectively. Both models are trained over a fixed number of epochs per dataset and then generate 10,000 molecules. 

As shown in Figure \ref{efficient}(a), OrgMol-Design exhibits substantial speed and graphics memory efficiency gains across all datasets during diffusion training. This improvement is most pronounced for datasets containing larger macromolecules, such as HCE, SNB-60K, and DTP, with time reduced by 4-7 times and memory by 3-4 times. For datasets like GDB-13, which primarily consist of smaller molecules, the efficiency improvement is more modest, with a reduction of about 2 times in both metrics. During molecular sampling (Figure \ref{efficient}(b)), OrgMol-Design also shows enhanced efficiency, especially in memory usage on the DTP dataset. These results suggest that OrgMol-Design's fragment-prior-based framework offers substantial efficiency gains over atom-level diffusion models like GDSS.

\section{Conclusion}

We propose OrgMol-Design, a novel multi-granularity framework for efficiently designing complex organics, which is composed of a score-based generative model via fragment prior for diverse coarse-grained scaffold generation and a chemical-rule-aware scoring model for fine-grained molecular structure design. Experimental results on four real-world benchmarks demonstrate the superiority and efficiency of our model in generating complex organic molecules.

\clearpage

\bibliography{aaai25}

\clearpage
\newpage
\appendix
\onecolumn

\section{Appendix}

\subsection{A\quad Fragment Vocabulary Construction}

Our goal is to construct a frequency-based molecular fragment vocabulary for a given dataset, drawing inspiration from prior works such as \cite{BPE,kong2022molecule}. The pseudo code outlining this process is provided in Algorithm \ref{vocab}. $\mathrm{Merge}(\cdot)$ represents combining neighboring fragments to generate new substructures and then converting them into corresponding SMILES representations. $\mathrm{Update}(\cdot)$ combines the fragments corresponding to a given SMILES string representing a potential new fragment formed by merging two existing fragments. It modifies the internal representation of the molecule by uniting the atom indices of the merged fragments, updating their SMILES, and removing the individual fragments that have been merged.

\begin{algorithm}[h]
\caption{Fragment Vocabulary Construction}
\label{vocab}
\begin{algorithmic}[1]
\Statex \hspace{-\algorithmicindent} \textbf{Input:} A molecule set $\mathcal{D} = \{\mathcal{G}_i = (\mathcal{V}_i, \mathcal{E}_i)\}_{i=1}^m$, desired vocabulary size $K$ 
\Statex \hspace{-\algorithmicindent} \textbf{Output:} A molecular fragment vocabulary $\mathbb{V}$
\Statex 
\State $\mathbb{V} \gets \{v\}$ ; \Comment{Initialize a vocabulary to all atoms $v$ in $\mathcal{D}$}

\For{$k = 1$ to $K - |\mathbb{V}|$}

    \State $\mathcal{C} \gets \{\}$ ; \Comment{Initialize an empty frequency counter}
    
    \For{$\mathcal{G}_d$ in $\mathcal{D}$}
        \For{$(\mathcal{F}_i,\mathcal{F}_j,\mathcal{E}_{ij})$ in $\mathcal{G}_d$}
            \State $\mathcal{F} \gets \mathrm{Merge}(\mathcal{F}_i,\mathcal{F}_j,\mathcal{E}_{ij})$ ; \Comment{Merge neighboring fragments into a novel fragment}
            \State $\mathcal{C}[\mathcal{F}] \gets \mathcal{C}[\mathcal{F}] + 1$ ; \Comment{Update the frequency of the fragment (initial value is $0$)}
        \EndFor
    \EndFor
    \State $\mathcal{F}_t \gets \arg\max_{(\mathcal{F}, f) \in \mathcal{C}} f$ ; \Comment{Identify the most frequent merged fragment}
    \State $\mathbb{V} \gets \mathbb{V} \cup \{\mathcal{F}_t\}$
    \State $\mathcal{D}' \gets \{\}$
    \For{$\mathcal{G}_d$ in $\mathcal{D}$}
        \State $\mathcal{G}'_d \gets \mathrm{Update}(\mathcal{G}_d,\mathcal{F}_t)$ ; \Comment{Update the molecules by merging all $\mathcal{F}_t$ fragments}
        \State $\mathcal{D}' \gets \mathcal{D}' \cup \{\mathcal{G}'_d\}$
    \EndFor
    \State $\mathcal{D} \gets \mathcal{D}'$
\EndFor
\State \textbf{return} $\mathbb{V}$
\end{algorithmic}
\end{algorithm}

\begin{figure}[h]
    \setlength{\abovecaptionskip}{0pt}
    \centering
    \includegraphics[width=0.65\linewidth]{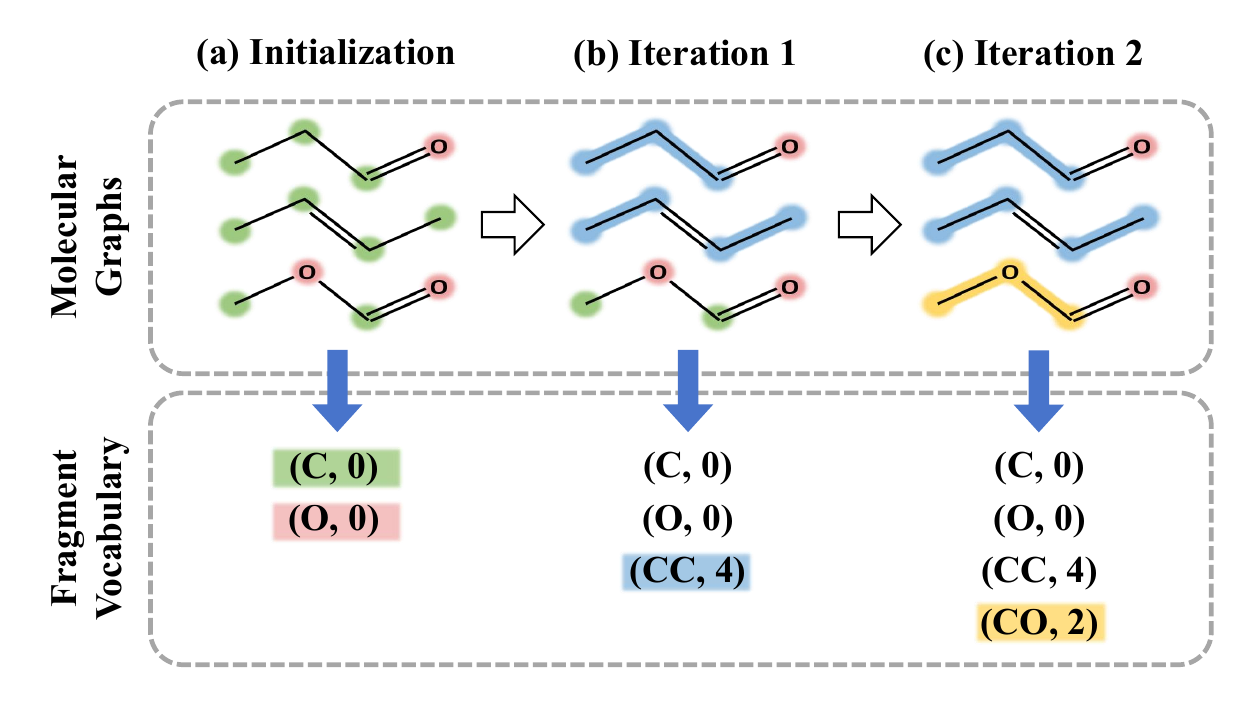}
    \caption{An example of fragment vocabulary construction on a given training set \{CCC=O, CC=CC, COC=O\}. (a) The vocabulary is initialized by single atoms. (b) In the first iteration, the fragment CC (highlighted in blue) emerges as the most frequent and is subsequently added to the vocabulary. All occurrences of CC are then merged to update the molecular graphs. (c) In the second iteration, CO (highlighted in yellow) is the most frequent, leading to its addition to the vocabulary and subsequent merging. After two iterations, the vocabulary is constructed as \{C, O, CC, CO\}.}
    \label{fragment}
\end{figure}

To further elucidate this process, we also provide an illustrative example of this fragment vocabulary construction process in Figure~\ref{fragment}.

\subsection{B\quad Coarse-grained Score-based Generative Modeling}

In this section, we provide the details of our proposed coarse-grained score-based generative model. 

\subsubsection{B.1\quad VE and VP SDEs}

We provide two types of stochastic differential equations (SDEs), i.e., the Variance Exploding (VE) SDE and Variance Preserving (VP) SDE, whose discretizations cause noise perturbations of our coarse-grained score-based generative model \cite{jo2022GDSS}.

The formulation of the VE SDE is given by:
\begin{equation}
    \mathrm{d}\mathbf{x} = \sigma_{min}\left(\frac{\sigma_{max}}{\sigma_{min}}\right)^{t} \sqrt{2\log\frac{\sigma_{max}}{\sigma_{min}}}\mathrm{d}\mathbf{w}, \,\, t \in (0,1],
    \label{VE}
\end{equation}
where $\sigma_{min}$ and $\sigma_{max}$ are predefined hyperparameters (detailed in Appendix H.2). The corresponding perturbation kernel is formulated as follows:
\begin{equation}
    p_{0t}(\mathbf{x}(t) \mid \mathbf{x}(0)) = \mathcal{N}\left(\mathbf{x}(t) \mid \mathbf{x}(0), \sigma_{min}^2 \Big( \frac{\sigma_{max}}{\sigma_{min}}\Big)^{2t} \mathbf{I}\right), \,\, t \in (0,1].
    \label{kernelVE}
\end{equation}

The process of VP SDE is as follows:
\begin{equation}
    \mathrm{d}\mathbf{x} = -\frac{1}{2}\beta_t\mathbf{x}\mathrm{d}t + \sqrt{\beta_t}\mathrm{d}\mathbf{w}, \,\, t \in (0,1],
    \label{VP}
\end{equation}
where $\beta_t=\beta_{min} + t(\beta_{max}-\beta_{min})$ with both $\beta_{max}$ and $\beta_{min}$ serve as hyperparameters (detailed in Appendix H.2). Accordingly, the perturbation kernel is expressed as:
\begin{equation}
\begin{aligned}
    p_{0t}(\mathbf{x}(t) \mid \mathbf{x}(0)) = \mathcal{N}\left(\mathbf{x}(t) \mid e^{-\frac{1}{4}t^2(\beta_{max}-\beta_{min}) - \frac{1}{2}t\beta_{min}}\mathbf{x}(0), \mathbf{I} - \mathbf{I} e^{-\frac{1}{2}t^2(\beta_{max} -\beta_{min}) - t\beta_{min}} \right), \,\, t \in (0,1].
\end{aligned}
\label{kernelVP}
\end{equation}

\begin{algorithm}[h]
\caption{Fine-grained Molecular Structure Design via Bond Scoring}
\label{bond_code}
\begin{algorithmic}[1]
\Statex \hspace{-\algorithmicindent}  \textbf{Input:} A molecular fragment set $\{\mathcal{F}_i = \{\tilde{\mathcal{V}}_i, \tilde{\mathcal{E}}_i\}\}_{i=1}^{m}$, a fragment-level adjacency matrix $\textbf{\em C} \in \mathbb{R}^{m \times m}$, the bond scoring model $\textbf{\em p}$, a mapping from each atom to its fragment $\boldsymbol{\omega}$, and a score threshold $\Psi_{th}$
\Statex \hspace{-\algorithmicindent} \textbf{Output:} A complete, valid molecular graph $\mathcal{G}$
\Statex
\State $\mathcal{G} \leftarrow \mathrm{Chem.RWMol()}$ \Comment{\textit{Create an empty molecular graph}}
\State $\mathcal{E}^{intra} \leftarrow \{\}$ \Comment{\textit{Create the intra-fragment bond set}}
\State $\mathcal{B} \leftarrow \{\}$ \Comment{\textit{Create the candidate inter-fragment bond set}}

\For {$\mathcal{F}_i$ in $\{\mathcal{F}_i\}_{i=1}^{m}$}
    \State $\mathcal{G} \leftarrow \mathrm{AddAtom}(\tilde{\mathcal{V}}_i)$ \Comment{\textit{Add intra-fragment atoms to $\mathcal{G}$}}
    \State $\mathcal{G} \leftarrow \mathrm{AddBond}(\tilde{\mathcal{E}}_i)$ \Comment{\textit{Add intra-fragment bonds to $\mathcal{G}$}}
    \State $\mathcal{E}^{intra} \leftarrow \mathcal{E}^{intra} \cup \tilde{\mathcal{E}}_i$ \Comment{\textit{Update the intra-fragment bond set}}
\EndFor

\For{$(u,v)$ where $u \in \mathcal{F}_i$ and $v \in \mathcal{F}_j$}
    \If{$(u,v) \notin \mathcal{E}^{intra}$ and $\textbf{\em C}(\boldsymbol{\omega}(u),\boldsymbol{\omega}(v))=1$}
        \State Compute the bond score $\mathcal{J}(u,v)$ via $\textbf{\em p}$
        \State $\mathcal{B} \leftarrow \mathcal{B} \cup \{(u,v,\mathcal{J}(u,v))\}$ \Comment{\textit{Add bond and score to the inter-fragment bond set}}
    \EndIf
\EndFor

\State $\mathcal{B} \leftarrow \mathrm{SortByScore}(\mathcal{B})$ \Comment{\textit{Sort the candidate bonds based on their scores}}

\For{$b$ in $\mathcal{B}$}
    \If{$\mathcal{J}(b) > \Psi_{th}$ and $\mathrm{ChemicalRuleCheck}(b)$}
        \State $\mathcal{G} \leftarrow \mathrm{AddBond}(b)$ \Comment{Incorporate a checked inter-fragment bond into $\mathcal{G}$}
    \EndIf
\EndFor

\State $\mathcal{G} \leftarrow \mathrm{LargestConnectedComponent}(\mathcal{G})$ \Comment{\textit{Determine the largest component as final molecule}}

\State \textbf{return} $\mathcal{G}$
\end{algorithmic}
\end{algorithm}

\subsubsection{B.2\quad Reverse Diffusion-driven Molecular Fragment Generation and Quantization}

To generate molecular fragments through the reverse diffusion process, we first sample $N$, representing the maximum number of fragments in a molecule, according to the empirical distribution of fragment counts observed in the training dataset. Subsequently, we sample noise with a batch size $B$ from the prior distribution. In this context, $\textbf{\em F}_T \in \mathbb{R}^{N \times K \times B}$ corresponds to fragment features, while $\textbf{\em C}_T \in \mathbb{R}^{N \times N \times B}$ captures inter-fragment connections, where $F$ is the molecular fragment vocabulary size. The reverse-time SDE process is then simulated to produce the final fragment features, $\textbf{\em F}_0$, and their corresponding connections, $\textbf{\em C}_0$. These outputs are then quantized to yield discrete fragments and their associated connections. Specifically, we determine the index of the maximum value along the second dimension of $\textbf{\em F}_0$ as the corresponding fragment. Furthermore, the entries of $\textbf{\em C}_0$ are quantized to $\{0,1\}$ with values in $(0,0.5)$ set to $0$, and those in $[0.5,1)$ set to $1$, indicating the absence or existence of a connection, respectively. The hyperparameters related to this process are detailed in Appendix H.

\subsection{C\quad Fine-grained Molecular Structure Design via Bond Scoring}

Algorithm \ref{bond_code} provides the pseudo code for our fine-grained fragment assembly strategy based on bond scoring. Initially, we employ RDKit \cite{Landrum2016RDKit2016_09_4} to add the atoms and bonds of each molecular fragment to the molecule being constructed, and record the edges within the fragment. Subsequently, for pairs of nodes residing in distinct fragments where a connection between the fragments is known, we include the corresponding edges to the candidate inter-fragment bond set and calculate their associated scores. The candidate edges are then sorted in descending order according to their scores. During iteration, if an edge has a score greater than the threshold and passes the chemical-rule check, it is added to the molecule. The chemical-rule check ensures adherence to valence rules and prevents the formation of unstable cycles consisting of fewer than five or more than six nodes. Given the possibility of generating disconnected graphs during this procedure, we select the largest connected component as the final molecular structure.

\subsection{D\quad Ablation Studies}

To verify the influence of different modules of our model on the quality of the generated molecules, we make variants of the original design. In the first variant, fragment-level connections $\textbf{\em C}$ are removed, limiting the fine-grained atom-level bond prediction to only consider nodes that do not belong to the same fragment. The second variant excludes the bond scoring model, which means randomly connecting the generated fragments at the atomic level. We use \textbf{Uniqueness}, \textbf{FCD}, and \textbf{Novelty} as the evaluation metrics. High uniqueness indicates greater diversity in the generated molecules, while high novelty reflects the generation of new, previously unseen molecules. A lower FCD score signifies that the distribution of the generated molecules more closely aligns with that of the training set in chemical space. Table \ref{Ablation} presents a comparison of our model's performance against these two variants, with the results representing the mean values derived from 3 independent runs.

We can conclude from Table \ref{Ablation} that removing the fragment-level connections significantly reduces the diversity of generated molecules, as indicated by the drop in uniqueness, and slightly worsens the alignment with the training set distribution, as reflected by the increase in FCD. However, the most substantial impact is observed when the fine-grained bond scoring model is removed. Without it, the model's ability to generate diverse, novel, and chemically realistic molecules is severely compromised, leading to a drastic drop in uniqueness and a significant increase in FCD. These results clearly demonstrate that both components are essential for maintaining the high quality of molecule generation, with the fine-grained bond scoring being particularly critical. Therefore, the design of our model, which integrates these two components, is necessary to ensure optimal performance across different datasets.

\begin{table}[h]
\fontsize{9}{11}\selectfont 
\setlength{\tabcolsep}{4pt}
\begin{center}
\scalebox{1.0}{\begin{tabular}{lcccccccc}
\toprule
\multirow{2}{*}{} & \multicolumn{3}{c}{\textbf{GDB-13}}               &  & \multicolumn{3}{c}{\textbf{SNB-60K}}                          \\ \cline{2-4} \cline{6-8} 
                        & Uniqueness(\%)            & FCD            & Novelty(\%)            &  & Uniqueness(\%)            & FCD           & Novelty(\%)            \\ \midrule
Ours                     & \textbf{99.36}          & \textbf{8.90}          & \textbf{99.14}           &  & \textbf{84.72}         & \textbf{5.19}         & \textbf{99.43}         \\
Ours w/o $\textbf{\em C}$                   & 96.77          & 9.20          & 98.77          &  & 77.21& 6.65& 99.40 \\
Ours w/o bond scoring                   & 1.92          & 22.32          & 89.58          &  &0.96 &14.72 & 82.25         \\
\bottomrule
\end{tabular}}
\caption{Comparative results of OrgMol-Design and its variants.}
\label{Ablation}
\end{center}
\end{table}

\subsection{E\quad Analysis of Fragment Proportions with Different Vocabulary Sizes}

\begin{figure}[h]
    \setlength{\abovecaptionskip}{0pt}
    \centering
    \includegraphics[width=0.9\linewidth]{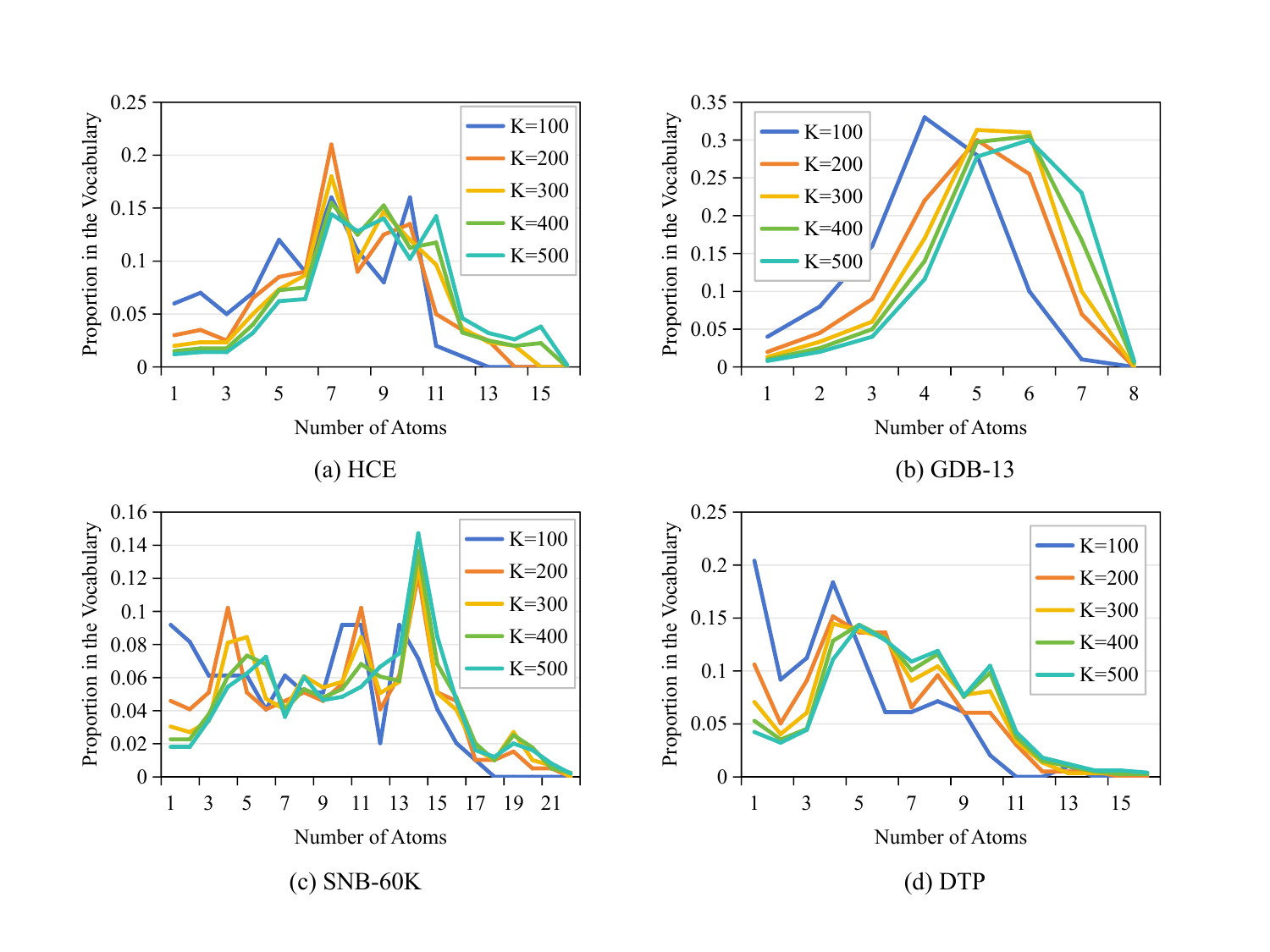}
    \caption{The proportional distribution of fragments with different atomic numbers across different vocabulary sizes in four datasets.}
    \label{Proportion}
\end{figure}

We investigate the distribution of fragment proportions categorized by their atomic counts across four datasets: HCE, GDB-13, SNB-60K, and DTP, under varying vocabulary sizes ($K=100,200,300,400,500$), as depicted in Figure \ref{Proportion}. 

Across all datasets, fragments consisting of 2 to 7 atoms dominate the vocabulary, with certain trends emerging based on the dataset characteristics. In the HCE and GDB-13 datasets, fragments with 5 to 7 atoms exhibit the highest proportions, suggesting a high prevalence of these atomic counts in the underlying molecular structures. In contrast, the SNB-60K dataset displays a more diversified distribution, with substantial proportions in both smaller fragments (3-5 atoms) and larger fragments (8-17 atoms). For the DTP dataset, the distribution follows a similar pattern to HCE, albeit with a more even spread across the range of atomic counts.

As the vocabulary size $K$ increases, the proportional distribution tends to stabilize, especially for the most frequently occurring fragment sizes, which implies that a larger vocabulary captures more diverse molecular structures without drastically altering the relative proportions of common occurring fragment sizes.

\subsection{F\quad Performance Analysis on Standard Evaluation Metrics}

In this section, we present the performance of our model using standard evaluation metrics. We assess the quality of 10,000 molecules generated by OrgMol-Design and GDSS based on the following metrics. \textbf{Validity} measures the percentage of chemically valid molecules. \textbf{Uniqueness} indicates the percentage of unique molecules. \textbf{Fréchet ChemNet Distance (FCD)} \cite{doi:10.1021/acs.jcim.8b00234} quantifies the distance between the reference and generated molecule set by comparing the activations of the final layer in ChemNet. The lower the FCD value, the higher the similarity in chemical space between two distributions. Finally, \textbf{Novelty} reflects the percentage of novel-generated molecules with reference to the training set.

\begin{figure}[h]
    \setlength{\abovecaptionskip}{0pt}
    \centering
    \includegraphics[width=0.9\linewidth]{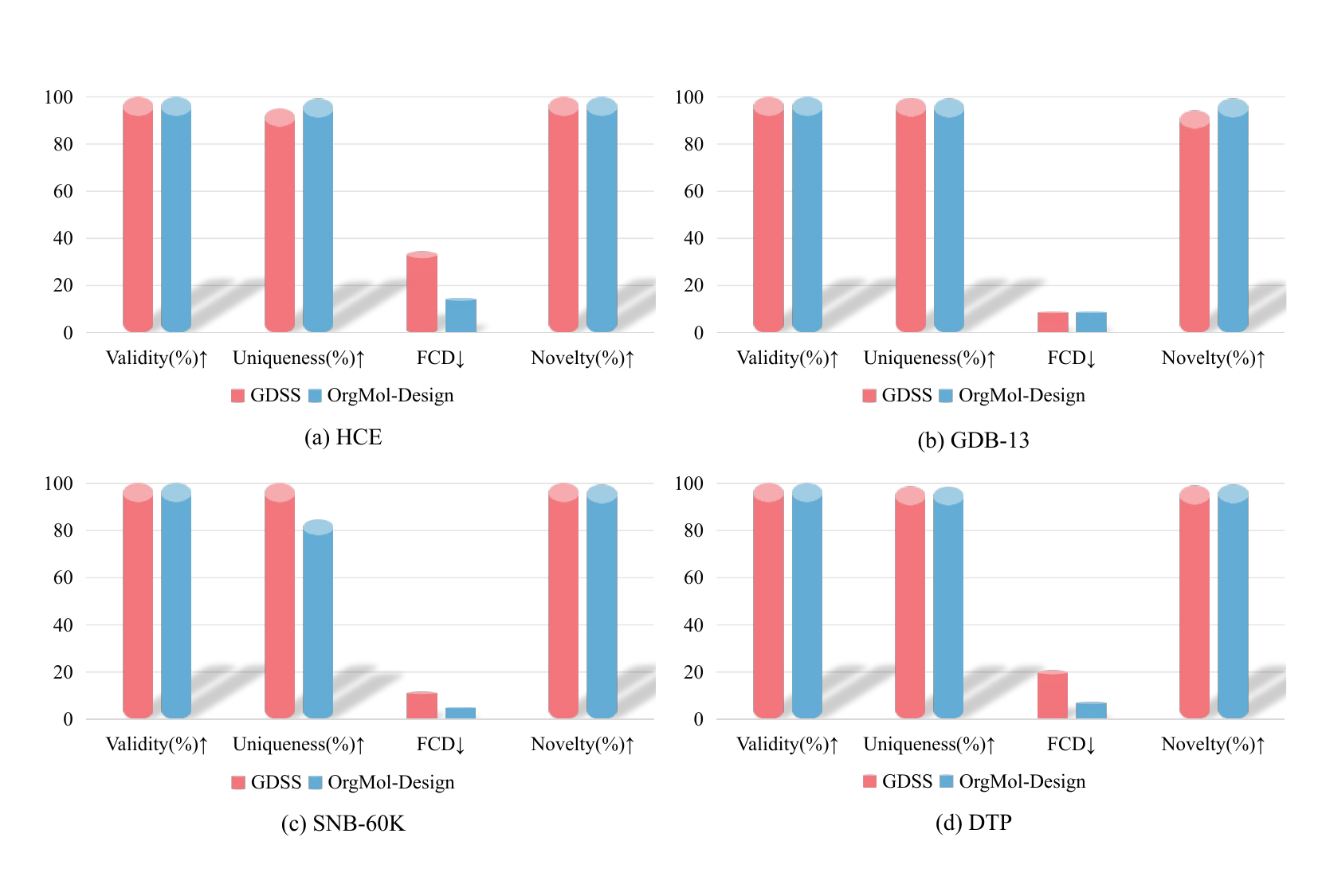}
    \caption{Comparative Performance of OrgMol-Design and GDSS on standard evaluation metrics across four datasets.}
    \label{Standard}
\end{figure}

We run three times and calculate the average, and the results are summarized in Figure \ref{Standard}. Across all datasets, both models achieve near-perfect scores in validity and uniqueness, indicating that the molecules generated by both OrgMol-Design and GDSS are predominantly chemically valid and unique. However, OrgMol-Design consistently outperforms GDSS in terms of FCD across all datasets. The lower FCD values suggest that OrgMol-Design generates molecules that are more chemically similar to the reference distributions, demonstrating the superior performance of our fragment-based mechanism in chemical space exploration. Furthermore, for the novelty metric, both models show high performance, though there are slight variations across different datasets. The results indicate that OrgMol-Design is more effective at generating novel molecules, further showcasing its capability to discover new chemical structures.

\subsection{G\quad Analysis of Atom and Fragment Distribution in Molecules}

\begin{figure}[h]
    \centering
    \includegraphics[width=0.8\linewidth]{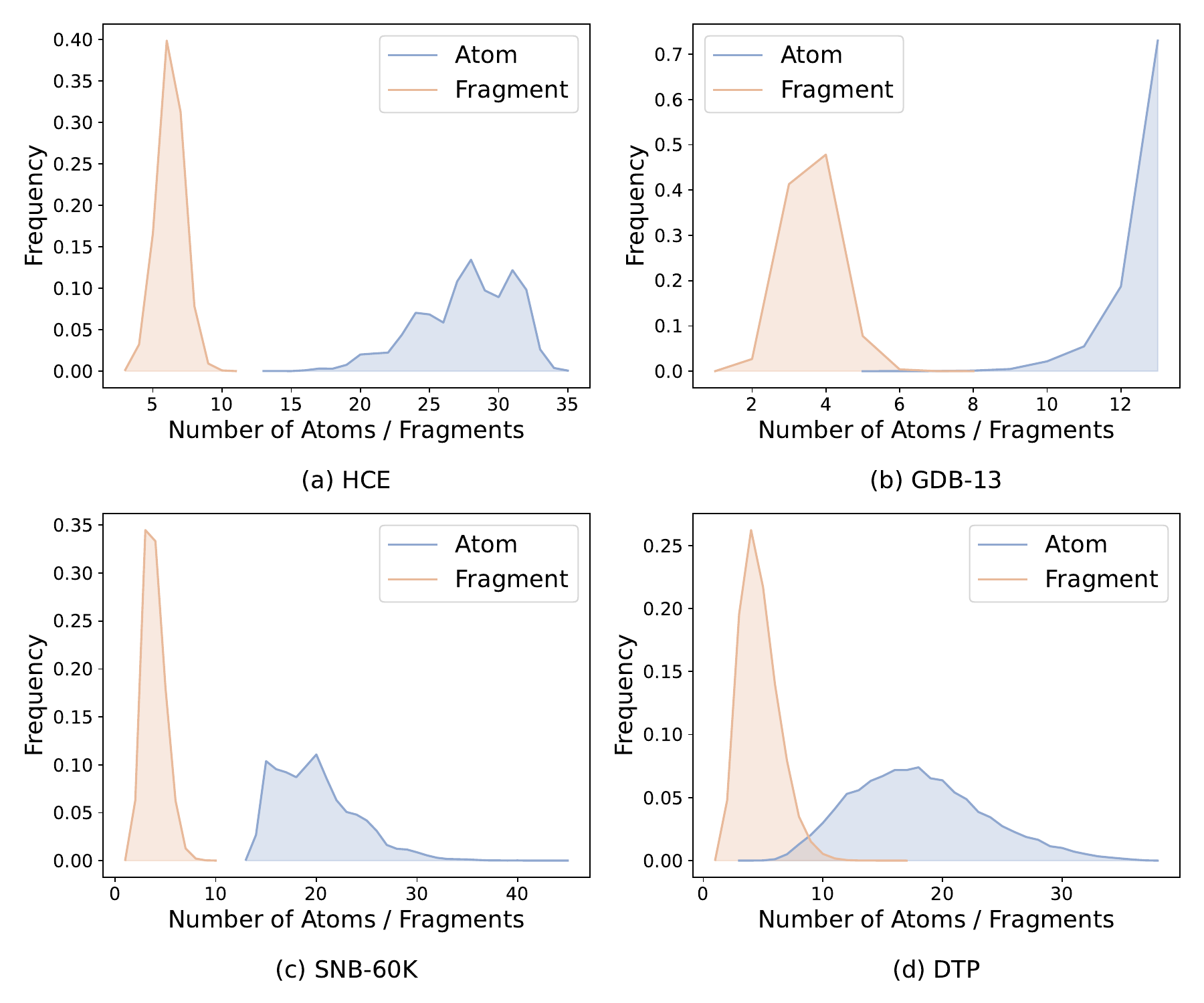}
    \caption{The distribution of the number of atoms and fragments in a molecule across four datasets.}
    \label{atom_fragment}
\end{figure}

In this section, we conduct an analysis of the frequency distributions of the number of atoms and fragments within each molecule across four distinct datasets (see Appendix H.1 for fragmentation details). These distributions are visualized in Figure \ref{atom_fragment}, which depicts the relative frequency of molecules with varying atom and fragment counts. 

We can observe that the majority of molecules in these datasets contain a moderate number of atoms, while the distribution of fragments is more intense. Across all four datasets, the number of fragments is mainly concentrated within 10, which is much smaller than the number of atoms. This finding underscores the potential of using fragments as fundamental units of description, which can considerably reduce the dimensionality of the model's input space, thereby enhancing computational efficiency. This result further validates the effectiveness of our model in capturing essential molecular characteristics.

\subsection{H\quad Hyperparameters}

\subsubsection{H.1\quad Hyperparameters of Fragmentation}

For the four datasets HCE, GDB-13, SNB-60K, and DTP, we set the respective molecular fragment vocabulary sizes to 100, 200, 100, and 200.

\subsubsection{H.2\quad Hyperparameters of Coarse-grained Score-based Generation}

The hyperparameters used in the coarse-grained score-based generation model across the four datasets are summarized in Table \ref{hp_coarse}, including the hyperparameters of the data definitions determined by the fragmentation process, the neural networks $\boldsymbol{\epsilon}_{\theta}$ and $\boldsymbol{\epsilon}_{\phi}$ used for score matching, the diffusion processes (i.e., the SDEs for $\textbf{\em F}$ and $\textbf{\em C}$), the SDE solver, and the training procedure.

\begin{table*}[h]
\fontsize{9}{11}\selectfont 
\setlength{\tabcolsep}{3pt}
    \vspace{-0.1in}
    \centering
    \begin{tabular}{llcccccc}
    \toprule
        & \textbf{Param} & \textbf{HCE} & \textbf{GDB-13} & \textbf{SNB-60K} & \textbf{DTP}  \\
    \midrule
        \multirow{2}{*}{Data}
        & Maximum number of fragments & 11 & 8 & 10 & 17 \\
        & Fragment feature dimension & 100 & 200 & 100 & 200 \\
    \midrule
        \multirow{2}{*}{$\boldsymbol{\epsilon}_{\theta}$}
        & Number of GCN layers & 2 & 2 & 2 & 2 \\
        & Hidden dimension & 16 & 16 & 16 & 16 \\
    \midrule
        \multirow{6}{*}{$\boldsymbol{\epsilon}_{\phi}$}
        & Number of attention heads & 4 & 4 & 4 & 4 \\
        & Number of initial channels & 2 & 2 & 2 & 2 \\
        & Number of hidden channels & 8 & 8 & 8 & 8 \\
        & Number of final channels & 4 & 4 & 4 & 4 \\
        & Number of GCN layers & 2 & 6 & 2 & 6 \\
        & Hidden dimension & 16 & 16 & 16 & 16 \\
    \midrule
        \multirow{4}{*}{SDE for $\textbf{\em F}$}
        & Type & VP & VP & VP & VP \\
        & Number of sampling steps & 1000 & 1000 & 1000 & 1000 \\
        & $\beta_{min}$ & 0.1 & 0.1 & 0.1 & 0.1 \\
        & $\beta_{max}$ & 1.0 & 1.0 & 1.0 & 1.0 \\
    \midrule
        \multirow{4}{*}{SDE for $\textbf{\em C}$}
        & Type & VE & VE & VE & VE \\
        & Number of sampling steps & 1000 & 1000 & 1000 & 1000 \\
        & $\beta_{min}$ & 0.1 & 0.2 & 0.1 & 0.2 \\
        & $\beta_{max}$ & 1.0 & 1.0 & 1.0 & 1.0 \\
    \midrule
        \multirow{4}{*}{Solver}
        & Predictor & Reverse & Reverse  & Reverse & Reverse \\
        & Corrector & Langevin & Langevin  & Langevin & Langevin \\
        & SNR & 0.2 & 0.2 & 0.2 & 0.2 \\
        & Scale coefficient & 0.5 & 0.9 & 0.5 & 0.9 \\
    \midrule
        \multirow{7}{*}{Train} 
        & Optimizer & Adam & Adam & Adam & Adam \\
        & Learning rate & 5e-3 & 5e-3 & 5e-3 & 5e-3 \\
        & Weight decay & 1e-4 & 1e-4 & 1e-4 & 1e-4 & \\
        & Batch size & 2048 & 8192 & 1024 & 2048 \\
        & Number of epochs & 300 & 500 & 300 & 500 \\
        & EMA & 0.999 & 0.999 & 0.999 & 0.999 \\
        & Learning rate decay & 0.999 & 0.999 & 0.999 & 0.999 \\
    \bottomrule
    \end{tabular}
    \vspace{0.5cm}
    \caption{Hyperparameters for coarse-grained score-based generation.}
    \label{hp_coarse}
\end{table*}

\subsubsection{H.3\quad Hyperparameters of Fine-grained Bond Scoring Model}

Table \ref{hp_fine} provides the model and training hyperparameters for the fine-grained bond scoring phase. In this model, each node feature is represented by a combination of atom and fragment embeddings.

\begin{table*}[h]
\fontsize{9}{11}\selectfont 
\setlength{\tabcolsep}{3pt}
    \vspace{-0.1in}
    \centering
    \begin{tabular}{llcccccc}
    \toprule
        & \textbf{Param} & \textbf{HCE} & \textbf{GDB-13} & \textbf{SNB-60K} & \textbf{DTP}  \\
    \midrule
        \multirow{5}{*}{Model}
        & Dimension of atom embeddings & 50 & 50 & 50 & 50 \\
        & Dimension of fragment embeddings & 100 & 100 & 100 & 100 \\
        & Dimension of node representations & 300 & 300 & 300 & 300 \\
        & Dimension of graph embeddings & 400 & 400 & 400 & 400 \\
        & Number of iterations of GINE & 4 & 4 & 4 & 4 \\
    \midrule
        \multirow{4}{*}{Train} 
        & Optimizer & Adam & Adam & Adam & Adam \\
        & Learning rate & 1e-3 & 1e-3 & 1e-3 & 1e-3 \\
        & Number of epochs & 10 & 10 & 10 & 10 \\
        & Batch size & 32 & 32 & 32 & 32 \\
    \bottomrule
    \end{tabular}
    \vspace{0.5cm}
    \caption{Hyperparameters of the fine-grained bond scoring model.}
    \label{hp_fine}
\end{table*}

\subsection{I\quad Details of Benchmarks}

In this section, we first present a comprehensive introduction to the four datasets employed in our benchmark experiments in Appendix I.1. Subsequently, Appendix I.2 to I.5 elaborate on the design details for organic photovoltaics, chemical reaction substrates, organic emissive materials, and protein ligands, respectively. Finally, we provide the settings of the baseline models in Appendix I.6.

\subsubsection{I.1\quad Dataset Details}

The datasets utilized in our experiments, specifically HCE, SNB-60K, GDB-13, and DTP, are detailed as follows, with their statistical characteristics summarized in Table \ref{dataset}.

\begin{table}[h]
    \centering
    \fontsize{9}{11}\selectfont 
    \begin{tabular}{@{}lcccc@{}}
    \toprule
    \textbf{Datasets} & \textbf{Number of graphs} & \textbf{Number of nodes} & \textbf{Number of node types}\\ 
    \midrule
    HCE & 24,953 & 13 $ \leq \mid \mathcal{V} \mid \leq $ 35 & 6  \\
    SNB-60K & 60,828  & 13 $ \leq \mid \mathcal{V} \mid \leq $ 45 & 8  \\
    GDB-13 & 398,453  & 5 $ \leq \mid \mathcal{V} \mid \leq $ 13 & 4  \\
    DTP  & 105,338 & 3 $ \leq \mid \mathcal{V} \mid \leq $ 38 & 20  \\
    \bottomrule
    \end{tabular}
    \vspace{0.5cm}
    \caption{Statistics of the HCE, SNB-60K, GDB-13, and DTP datasets.}
    \label{dataset}
\end{table}

\begin{itemize}
\item[$\bullet$] \textbf{HCE}
\end{itemize}

HCE is a subset of the Harvard Clean Energy Project Database \cite{C3EE42756K}, which aims to discover organic molecular materials for organic photovoltaic applications through high-throughput virtual screening. This database captures crucial quantum chemical properties, including electronic properties, band gaps, molecular orbitals, and excited-state characteristics for each molecule. HCE is composed of approximately 25,000 molecules, each containing an average of 28 non-hydrogen atoms. The atomic types in HCE encompass carbon (C), nitrogen (N), oxygen (O), sulfur (S), selenium (Se), and silicon (Si).

\begin{itemize}
\item[$\bullet$] \textbf{SNB-60K}
\end{itemize}

SNB-60K is a dataset comprising approximately 60,000 molecules, all characterized by the presence of a syn-sesquinorbornene structural unit. This dataset presents a diverse range of molecular structures, with an average of 20 non-hydrogen atoms per molecule, rendering it a robust collection for chemical reaction substrate design studies. The atomic composition of SNB-60K includes bromine (Br), carbon (C), chlorine (Cl), fluorine (F), iodine (I), nitrogen (N), oxygen (O), and sulfur (S), reflecting the chemical diversity that can be explored within this collection.

\begin{itemize}
\item[$\bullet$] \textbf{GDB-13}
\end{itemize}

GDB-13 is a dataset comprising roughly 400,000 organic molecules, each containing no more than 13 non-hydrogen atoms \cite{gdb13}. The atomic types present in GDB-13 include carbon (C), nitrogen (N), oxygen (O), and sulfur (S). These molecules have undergone comprehensive filtering, including constraints implemented through RDKit \cite{Landrum2016RDKit2016_09_4}, to ensure they possess cyclic and highly conjugated structures, specifically targeting extended planar $\pi$-systems. These features enhance the representation of structurally relevant spaces for potential organic emitters. Additionally, the synthetic accessibility (SA) scores for these molecules do not exceed 4.5.

\begin{itemize}
\item[$\bullet$] \textbf{DTP}
\end{itemize}

DTP, derived from the Developmental Therapeutics Program Open Compound Collection \cite{NCIDatabases,doi:10.1021/ci010056s}, comprises approximately 100,000 molecules that have been evaluated for therapeutic potential in the treatment of cancer and acquired immunodeficiency syndrome (AIDS) \cite{doi:10.1021/ci00021a032}. These molecules adhere to rigorous structural constraints (the filter described above), with an average of 18 non-hydrogen atoms per molecule. The atomic composition of DTP spans a wide array of elements, including gallium (Ga), antimony (Sb), fluorine (F), bismuth (Bi), indium (In), selenium (Se), phosphorus (P), boron (B), germanium (Ge), bromine (Br), nitrogen (N), sulfur (S), chlorine (Cl), thallium (Tl), oxygen (O), tellurium (Te), mercury (Hg), arsenic (As), carbon (C), and lead (Pb).

\subsubsection{I.2\quad Design of Organic Photovoltaics}

Organic photovoltaics (OPVs) are pivotal in advancing renewable energy technologies by optimizing the efficiency, cost-effectiveness, and application flexibility of organic solar cells (OSCs). Despite substantial progress, OPVs still face challenges with lower power conversion efficiencies (PCEs), which are key for assessing the practicality and performance in solar energy conversion \cite{tartarus}. 

To address this, we introduce two benchmark tasks trained on the HCE dataset, aimed at discovering novel organic photoactive materials with superior PCEs. The first task involves designing a small organic donor molecule to pair with [6,6]-phenyl-C61-butyric acid methyl ester (PCBM) as the acceptor in a bulk heterojunction device \cite{hachmann2014lead}. The second task focuses on designing a small organic acceptor molecule for use in bulk heterojunction devices with poly[N-90-heptadecanyl-2,7-carbazole-alt-5,5-(40,70-di-2-thienyl-20,10,30-benzothiadiazole)] (PCDTBT) as the donor \cite{lopez2017design}. The objectives of the above two tasks are defined by the difference between the PCEs and the synthetic accessibility (SA) scores \cite{ertl2009estimation} of the corresponding molecular structures, as follows:

\begin{itemize}
\item[$\bullet$] Maximizing $PCE_{PCBM} - SAscore$ ;
\item[$\bullet$] Maximizing $PCE_{PCDTBT} - SAscore$ .
\end{itemize}

The simulation workflow for calculating PCEs begins by accepting a molecular input in the form of a SMILES string \cite{weininger1989}. Initial Cartesian coordinates are generated using Open Babel \cite{oboyleOpenBabelOpen2011}, which are then subjected to a conformer search conducted by CREST \cite{pracht2020conformer}. After conformer selection, geometry optimization is carried out utilizing the XTB method \cite{bannwarthGFN2xTBAccurateBroadly2019}. Following optimization, a single-point energy calculation at the GFN2-xTB level \cite{bannwarthGFN2xTBAccurateBroadly2019} is performed, which yields key electronic properties such as the energies of the highest occupied molecular orbital (HOMO), the lowest unoccupied molecular orbital (LUMO), the HOMO-LUMO energy gap, and the molecular dipole moment. These calculated properties are subsequently employed in the Scharber model \cite{ameri2009organic} to estimate the PCE of the organic photovoltaic device. This workflow integrates both quantum chemical calculations and theoretical performance prediction models to streamline the evaluation of OPV candidates.

\subsubsection{I.3\quad Design of Chemical Reaction Substrates}

The development of novel chemical reactions is a critical pursuit that drives innovation in drug and material discovery, as well as the advancement of sustainable production methodologies \cite{schwallerMachineIntelligenceChemical2022}. Through the application of transition state (TS) theory \cite{truhlar1996current}, basic reaction parameters such as thermodynamic feasibility, reaction rate, and selectivity can be calculated from first principles, which necessitates the explicit modeling of the corresponding TS. However, current computational algorithms for molecular simulations are plagued by high failure rates, limited robustness, and exceedingly high computational costs, leading to a scarcity of reliable organic design benchmarks related to chemical reactivity ~\cite{jorner2021machine}. To address these challenges and enhance the reliability and efficiency of TS modeling, we adopt the SEAM force field method \cite{jensenLocatingMinimaSeams1992}. This approach integrates two force fields that are directly linked via intrinsic reaction coordinates with the target TS, enabling the construction of an effective TS force field.

Leveraging the SEAM force field method, we model the intramolecular concerted double hydrogen transfer reaction of syn-sesquinorbornenes, with only one TS connecting the reactants and products \cite{reetz1972dyotropic}. We use this reaction to define a benchmark for modifying substrate and product structures to alter reactivity. We select activation energy and reaction energy as the primary properties for defining the reactivity of the system. The benchmark objectives for chemical reactivity are outlined as follows:

\begin{itemize}
\item[$\bullet$] Minimizing the activation energy $\Delta E_{a}$ ;
\item[$\bullet$] Minimizing the reaction energy $\Delta E_{r}$ ;
\item[$\bullet$] Minimizing the sum $\Delta E_{a}+\Delta E_{r}$ ;
\item[$\bullet$] Minimizing the difference $-\Delta E_{a}+\Delta E_{r}$ .
\end{itemize}

We perform this benchmark on the SNB-60K dataset, where each molecule contains a syn-sesquinorbornene structural unit. The simulation workflow begins with verification of hard constraints \cite{tartarus} in the SMILES string of the proposed substrate, ensuring that all generated molecules preserve the syn-sesquinorbornene core and remain stable. For the two objectives involving combinations of target properties, an additional constraint is imposed: the SAscore must not exceed six. Upon satisfying these constraints, initial Cartesian coordinates for the reactant and product are generated using the CREST conformer search \cite{pracht2020conformer}, initiated through Open Babel \cite{oboyleOpenBabelOpen2011}. Following this, the SEAM optimization process is conducted, where an initial geometry for the transition state is obtained by interpolating between the optimized geometries of the reactant and product (via polanyi optimization). The SEAM optimization then refines the guessed transition state structure, leading to the identification of the transition state, which is further optimized through constrained conformational sampling using CREST and finalized with polanyi optimization. In the final stage, the energy of the reactant ($E_{R}$), transition state ($E_{TS}$), and product ($E_{P}$) are calculated. From these, the reaction energy ($\Delta E_{r} = E_{P} - E_{R}$) and the approximate SEAM activation energy ($\Delta E_{a} = E_{TS} - E_{R}$) are extracted, providing critical insights into the energetic profile of the reaction.

\subsubsection{I.4\quad Design of Organic Emitters}

The design of organic emissive materials for organic light-emitting diodes (OLEDs) has garnered widespread attention in recent years, particularly following the discovery of thermally activated delayed fluorescence (TADF) \cite{uoyama2012highly}. These materials are primarily utilized in digital displays and lighting applications \cite{wong2017purely}. To enhance both the efficiency and longevity of OLED devices, TADF emitters are engineered to minimize the energy gap between the first excited singlet and triplet states, referred to as the singlet-triplet gap \cite{gomez2016design}. Furthermore, it is necessary to increase the fluorescence rate, which corresponds to maximizing the oscillator strength between the first excited singlet and the ground state \cite{gomez2016design}. The development of efficient blue emissive materials for OLEDs is especially challenging, requiring the design of molecules whose excitation energy between the ground state and the first excited singlet state corresponds to the energy of blue light \cite{wong2017purely,gomez2016design}. The objectives of these three design tasks are summarized as follows:

\begin{itemize}
\item[$\bullet$] Minimizing the singlet-triplet gap (ST) : $\Delta E(S_1$-$T_1)$ ;
\item[$\bullet$] Maximizing the oscillator strength (OSC) for the transition between $S_1$ and $S_0$ ;
\item[$\bullet$] Maximizing the combined objective : $+ OSC - ST - |\Delta E(S_0$-$S_1) - 3.2 \; eV|$ .
\end{itemize}

In this benchmark, high-fitness molecules must exhibit a SAscore of 4.5 or lower. The workflow initiates with the generation of a molecule, where the initial geometry is obtained through Open Babel \cite{oboyleOpenBabelOpen2011} and RDKit \cite{Landrum2016RDKit2016_09_4}. This geometry undergoes a conformer search using CREST \cite{pracht2020conformer} to explore possible low-energy conformations. Following the conformer search, the geometry is further optimized using the XTB method \cite{grimme2017a,bannwarthGFN2xTBAccurateBroadly2019} to refine the molecular structure. Subsequently, excited state properties are calculated via TD-DFT single-point calculations \cite{hariharan1973influence}, performed using the PySCF package \cite{sun2018pyscf}. This stage yields key electronic properties, including the singlet-triplet energy gap ($\Delta E(S_1$-$T_1)$), oscillator strength, and vertical excitation energy ($\Delta E(S_0$-$S_1)$). These properties are critical for evaluating the photophysical behavior of the organic emitter \cite{tartarus}.

\subsubsection{I.5\quad Design of Protein Ligands}

Designing small molecule ligands for specific proteins is a crucial endeavor in the advancement of targeted therapeutic strategies and molecular biology research. These ligands not only serve as potential drug candidates by modulating the activity of disease-related proteins, but also function as indispensable tools for probing protein function and interactions within biological systems. To this end, we develop a benchmark specifically aimed at the design of ligands for a specific protein based on molecular docking simulations. For this study, we select 4LDE, the $\beta$2-adrenoceptor GPCR receptor, which spans the cell membrane and binds adrenaline, a hormone responsible for mediating muscle relaxation and bronchodilation \cite{ring2013adrenaline}. The objectives of the benchmark are summarized as follows:

\begin{itemize}
\item[$\bullet$] Minimizing docking scores (DS) ;
\item[$\bullet$] Maximizing the success rate (SR) for sampled molecules passing structure filters.
\end{itemize}

Notably, the benchmark's objectives are not solely determined by docking scores but also incorporate stringent structural constraints \cite{tartarus}. If these constraints are not met, an exceedingly unfavorable score of 10,000 is assigned in place of the actual docking score. The list of these filters includes:

\begin{itemize}
\item[$\bullet$] Absence of reactive groups.
\item[$\bullet$] Absence of formal charges.
\item[$\bullet$] Absence of radicals.
\item[$\bullet$] At most 2 bridgehead atoms.
\item[$\bullet$] No rings larger than 8-membered.
\item[$\bullet$] Fulfils Lipinski's Rule of Five.
\item[$\bullet$] $SAscore < 4.5$.
\item[$\bullet$] $QED > 0.3$.
\item[$\bullet$] $TPSA > 140$.
\item[$\bullet$] Molecule passes the PAINS and WEHI and MCF filters. 
\item[$\bullet$] Molecule does not contain Si and Sn atoms.
\end{itemize}

This benchmark is conducted using the DTP dataset. The simulation workflow is initiated with the SMILES string of the proposed molecule, followed by the creation of an initial Cartesian coordinate guess using Open Babel \cite{oboyleOpenBabelOpen2011}, which serves as the starting structure for the docking procedure. The structure is then subjected to molecular docking using QuickVina2 \cite{alhossary2015fast} and Smina \cite{koes2013lessons}, both of which are molecular docking software tools. QuickVina2 focuses on speed and sampling efficiency, whereas Smina prioritizes accurate scoring and pose refinement. The final output is the docking score, which quantifies the molecule's binding energy to the target site, providing critical insights into the molecule's potential efficacy as a ligand for the specified protein target.

\subsubsection{I.6\quad Baseline Settings}

To evaluate the performance of our model on the four datasets mentioned above, we select SMILES-VAE \cite{doi:10.1021/acscentsci.7b00572}, SMILES-LSTM-HC \cite{doi:10.1021/acs.jcim.8b00839}, MoFlow \cite{10.1145/3394486.3403104}, REINVENT \cite{Olivecrona2017MolecularDD}, GB-GA \cite{C8SC05372C}, GDSS \cite{jo2022GDSS}, and MiCaM \cite{geng2023de} as the baseline models and report the mean and standard deviation of their optimal objective values over five independent experiments. Specifically, the parameters for SMILES-VAE, SMILES-LSTM-HC, MoFlow, REINVENT, and GB-GA adhere to the configurations outlined in \cite{tartarus}, while those for GDSS and MiCaM are based on the settings proposed in their respective original works.  

\subsection{J\quad Visualization of Molecular Fragments}

\begin{figure}[t]
    \centering
    \includegraphics[width=0.85\linewidth]{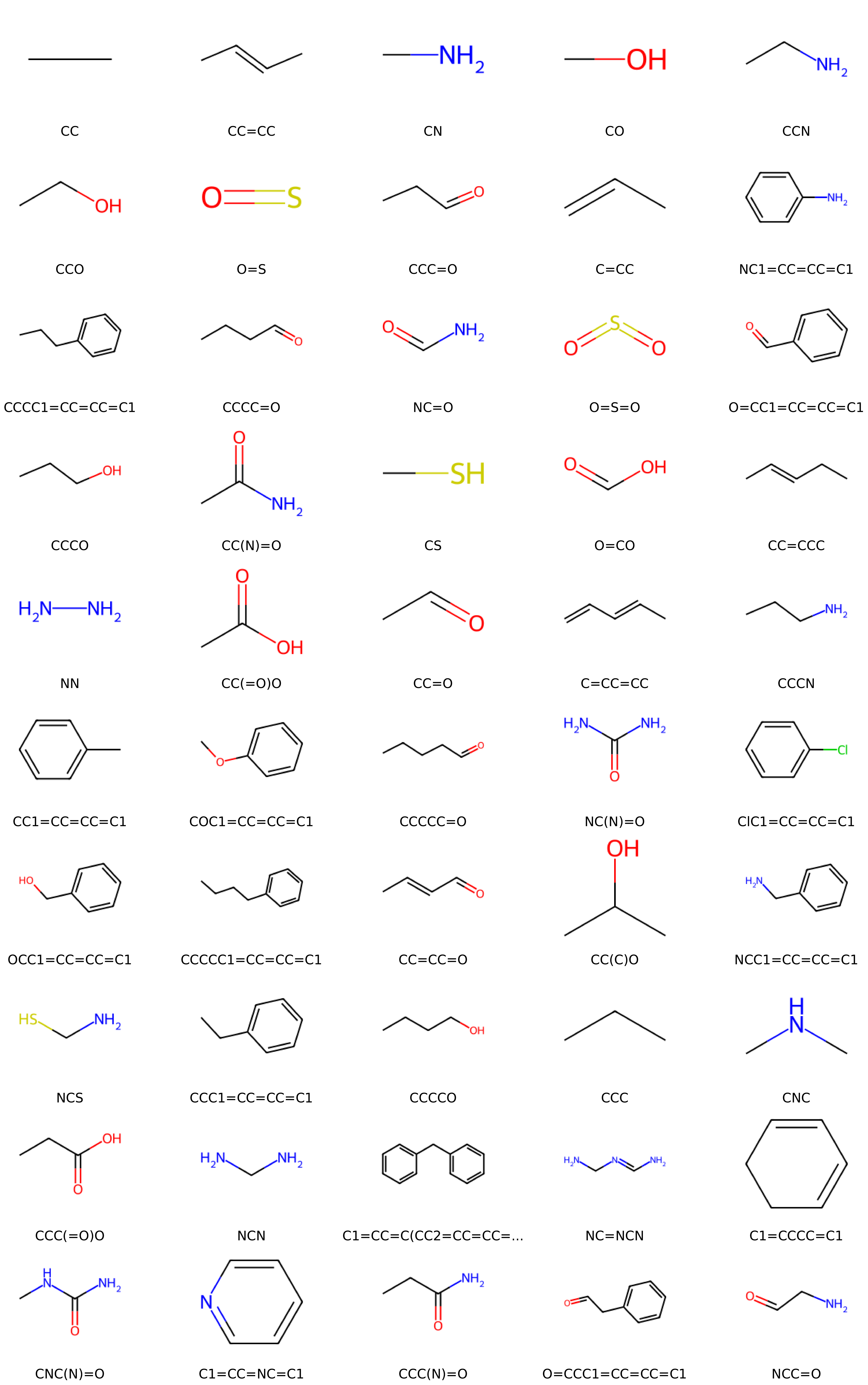}
    \caption{The top-50 fragments of the DTP dataset.}
    \label{DTP}
\end{figure}

\begin{figure}[t]
    \centering
    \includegraphics[width=0.85\linewidth]{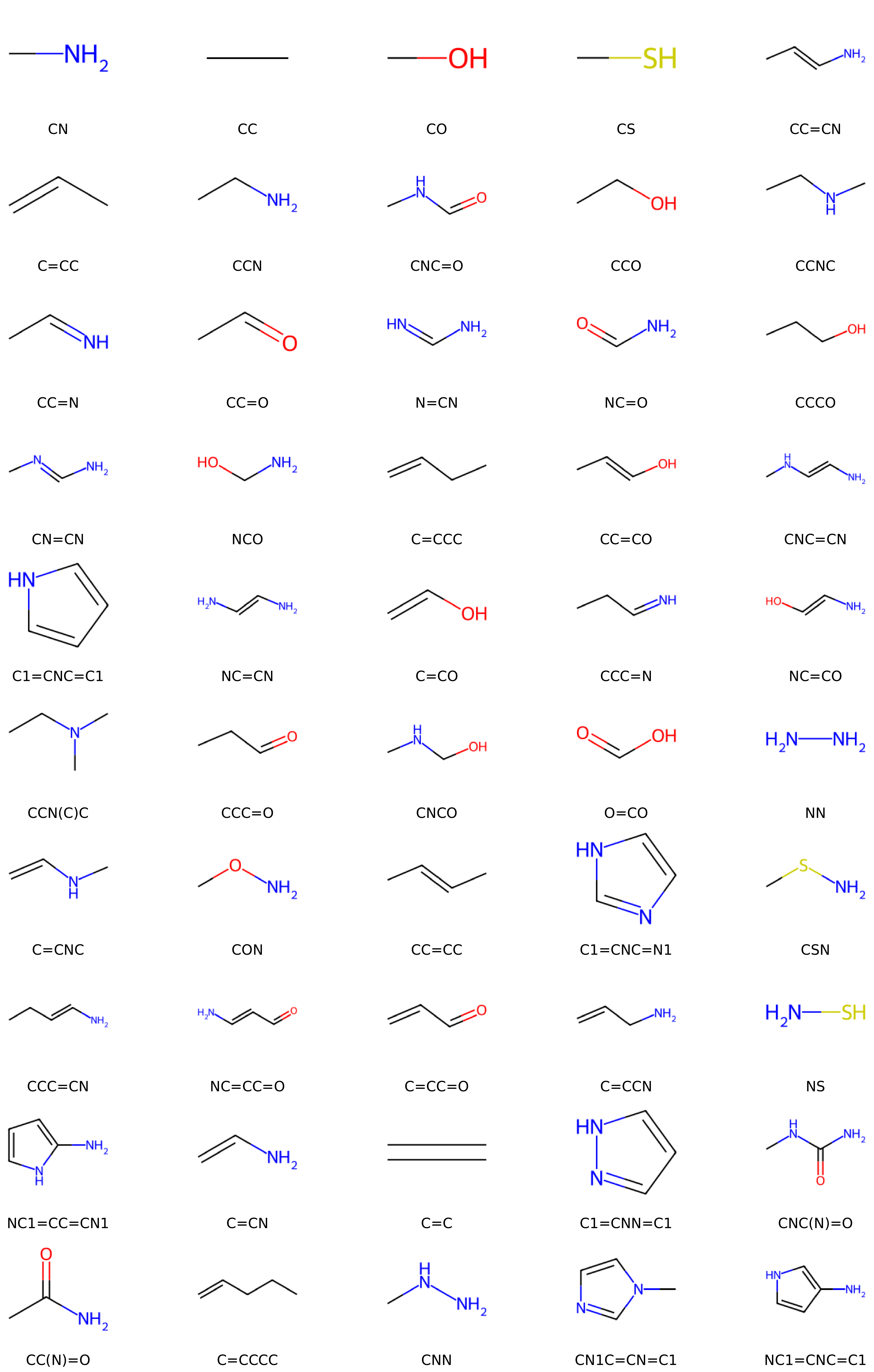}
    \caption{The top-50 fragments of the GDB-13 dataset.}
    \label{GDB-13}
\end{figure}

\begin{figure}[t]
    \centering
    \includegraphics[width=0.85\linewidth]{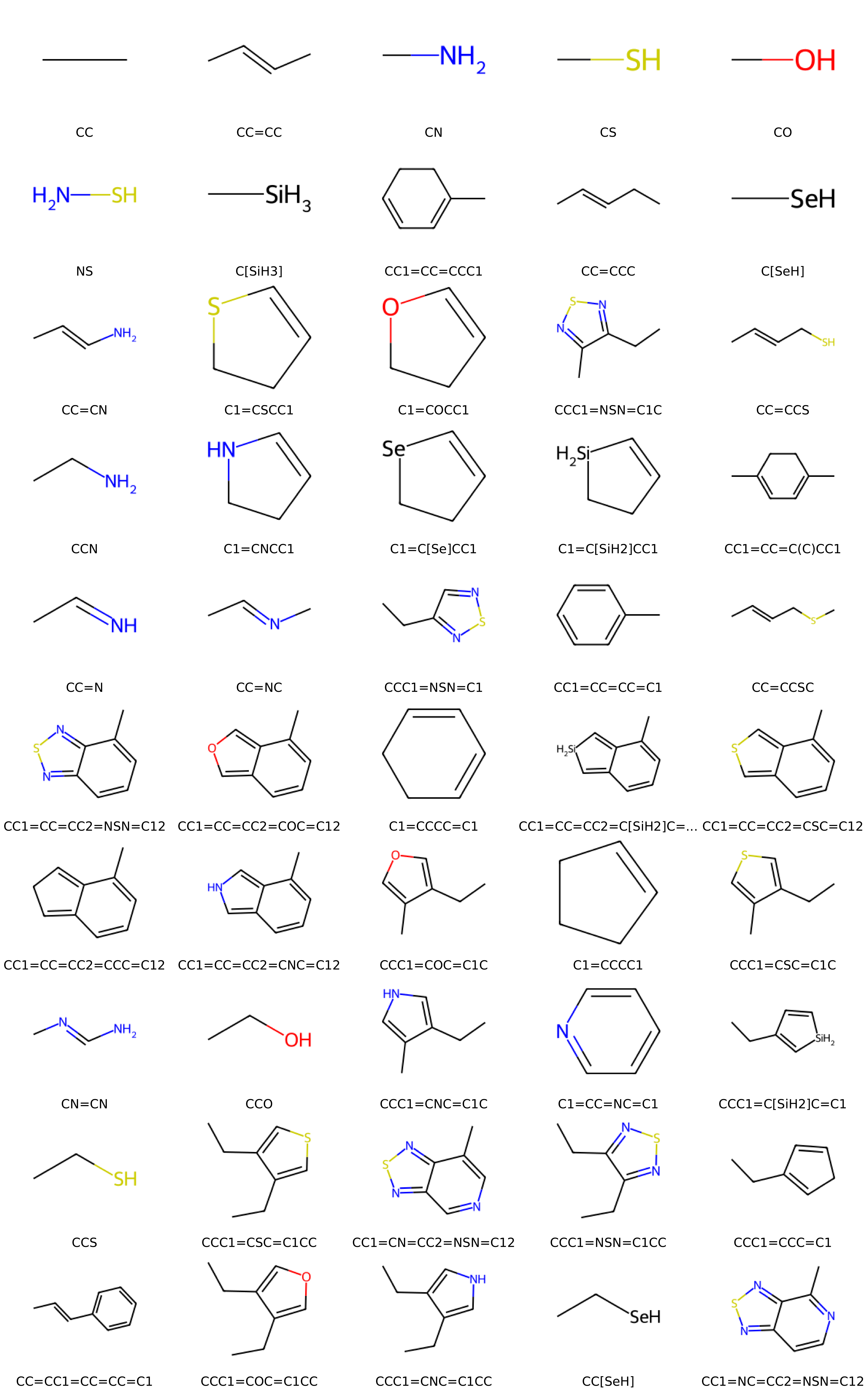}
    \caption{The top-50 fragments of the HCE dataset.}
    \label{HCE}
\end{figure}

\begin{figure}[t]
    \centering
    \includegraphics[width=0.85\linewidth]{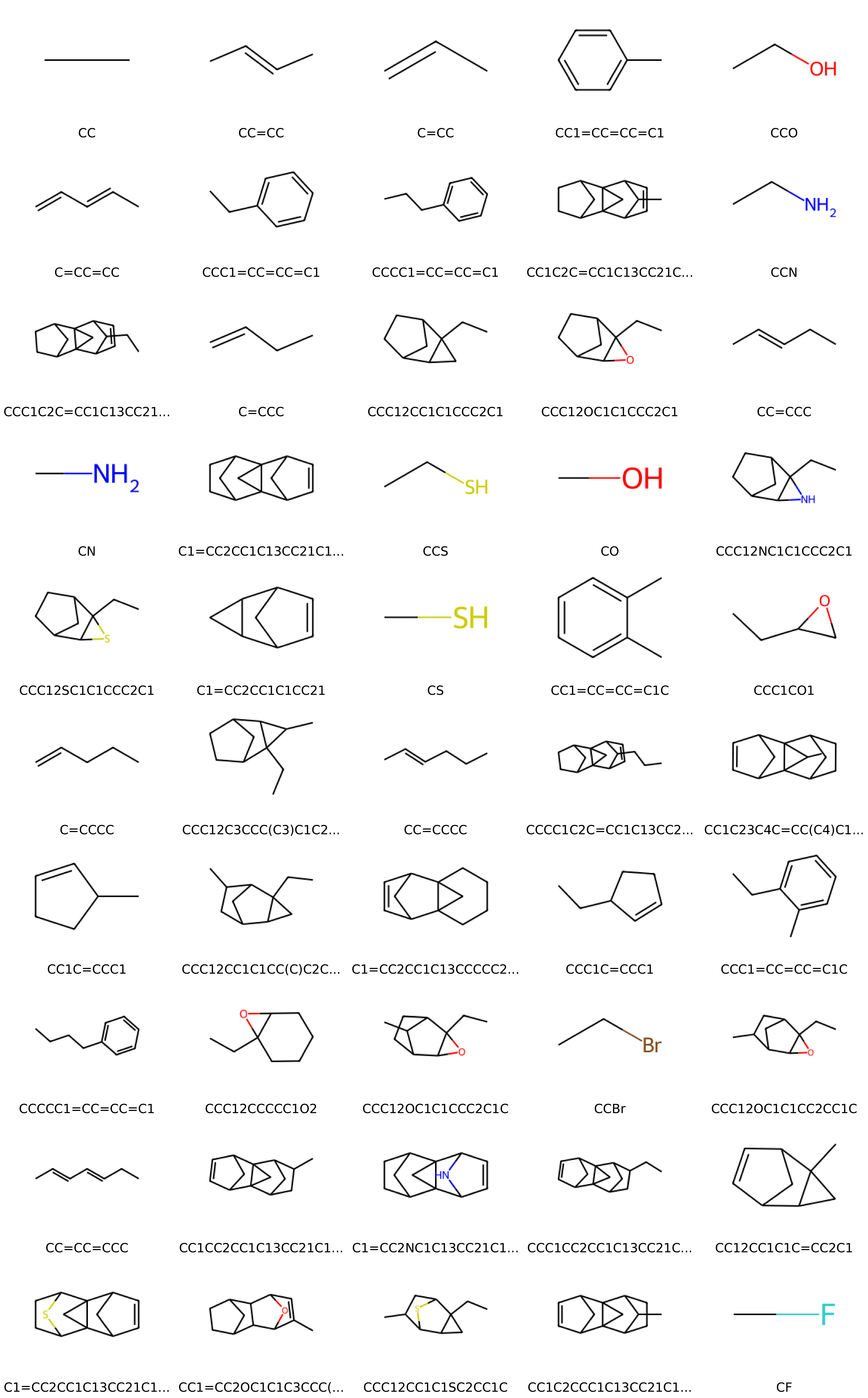}
    \caption{The top-50 fragments of the SNB-60K dataset.}
    \label{SNB-60K}
\end{figure}

We visualize the top 50 frequency-ranked fragments extracted from the DTP, GDB-13, HCE, and SNB-60K datasets using our fragmentation algorithm in Figure \ref{DTP}-\ref{SNB-60K}, respectively.


\end{document}